\documentclass[twocolumn]{aastex631}
\usepackage{float}
\usepackage{amsmath}	%
\usepackage{todonotes}

\newcommand{\deletedOne}[1]{}

\renewcommand{\Vec}[1]{{\bf #1}}

\newcommand*\cellwidth{\ensuremath{\Delta x}}

\newcommand*\xigo{\ensuremath{\xi_{\rm GO}}}
\newcommand*\xish{\ensuremath{\xi_{\rm sh}}}

\newcommand*\vwindhat{\ensuremath{\hat{\bf v}_{\bf {\rm wind}}}}

\newcommand*\vrel{\ensuremath{v_{\rm rel}}}
\newcommand*\vinflow{\ensuremath{v_{\rm inflow}}}
\newcommand*\vturb{\ensuremath{v_{\rm turb}}}
\newcommand*\Mturb{\ensuremath{\mathcal{M}_{\rm turb}}}

\newcommand*\lcool{\ensuremath{\ell_{\rm cool}}}

\newcommand*\logXe{\ensuremath{\log_\chi e/e_{\rm cl}}}

\newcommand*\rhocl{\ensuremath{\rho_{\rm cl}}}
\newcommand*\rhow{\ensuremath{\rho_{\rm w}}}
\newcommand*\rhomix{\ensuremath{\rho_{\rm mix}}}
\newcommand*\nmix{\ensuremath{n_{\rm mix}}}
\newcommand*\ncl{\ensuremath{n_{\rm cl}}}
\newcommand*\nw{\ensuremath{n_{\rm w}}}
\newcommand*\emix{\ensuremath{e_{\rm mix}}}
\newcommand*\ecl{\ensuremath{e_{\rm cl}}}
\newcommand*\ew{\ensuremath{e_{\rm w}}}
\newcommand*\ebreak{\ensuremath{e_{\rm break}}}
\newcommand*\emincool{\ensuremath{e_{{\rm min},{\rm cool}}}}
\newcommand*\Tmix{\ensuremath{T_{\rm mix}}}
\newcommand*\Tcl{\ensuremath{T_{\rm cl}}}
\newcommand*\Tw{\ensuremath{T_{\rm w}}}
\newcommand*\Tbreak{\ensuremath{T_{\rm break}}}
\newcommand*\rcl{\ensuremath{R_{\rm cl}}}
\newcommand*\rclcrit{\ensuremath{R_{\rm cl,crit}}}

\newcommand*\vw{\ensuremath{v_{\rm w}}}

\newcommand*\Mw{\ensuremath{\mathcal{M}_{\rm w}}}
\newcommand*\tshear{\ensuremath{t_{\rm shear}}}
\newcommand*\tcc{\ensuremath{t_{\rm cc}}}

\newcommand*\tcool{\ensuremath{t_{\rm cool}}}
\newcommand*\tcoolmix{\ensuremath{t_{\rm cool, mix}}}
\newcommand*\tcoolcl{\ensuremath{t_{\rm cool, cl}}}
\newcommand*\tcoolw{\ensuremath{t_{\rm cool, w}}}

\newcommand*\cs{\ensuremath{c_s}}
\newcommand*\cscold{\ensuremath{c_{s,{\rm cold}}}}
\newcommand*\csbreak{\ensuremath{c_{s,{\rm break}}}}
\newcommand*\cshot{\ensuremath{c_{s,{\rm hot}}}}

\newcommand*\tcoolmin{\ensuremath{t_{\rm cool, min}}}

\newcommand*\vrlike{\ensuremath{v_{r-{\rm like}}}}

\newcommand*\vphilike{\ensuremath{v_{\phi-{\rm like}}}}

\newcommand*\lturbsonic{\ensuremath{\ell_{\rm turb, sonic}}}

\newcommand*\KH{KH}
\newcommand*\pdns{principal dimensionless numbers}

\newcommand{\vsffirst}[1][\ell]{\ensuremath{\langle |\delta v|\rangle (#1)}}
\newcommand{\vsfsecond}[1][\ell]{\ensuremath{\langle (\delta v)^2\rangle (#1)}}

\newcommand*\fidtime{\ensuremath{2.5{\tcc}}}

\newcommand*\enzoe{{\sc enzo-e}}
\newcommand*\enzo{{\sc enzo}}
\newcommand*\cello{{\sc cello}}
\newcommand*\grackle{{\sc grackle}}

\newcommand*\citepAbruzzoSurvival{(Abruzzo et al., in prep.)}

\shorttitle{Taming the TuRMoiL}
\shortauthors{Abruzzo et al.}
\graphicspath{{./}{figures/}}

\begin{document}

\title{Taming the TuRMoiL: The Temperature Dependence of Turbulence in Cloud-Wind Interactions}

\correspondingauthor{Matthew W. Abruzzo}
\email{mwa2113@columbia.edu}

\author[0000-0002-7918-3086]{Matthew W. Abruzzo}
\affiliation{Department of Astronomy,
Columbia University, 550 West 120th Street,
New York, NY 10027, USA}

\author[0000-0003-3806-8548]{Drummond B. Fielding}
\affiliation{Center for Computational Astrophysics,
Flatiron Institute, 162 5th Avenue,
New York, NY 10010, USA}

\author[0000-0003-2630-9228]{Greg L. Bryan}
\affiliation{Department of Astronomy,
Columbia University, 550 West 120th Street,
New York, NY 10027, USA}
\affiliation{Center for Computational Astrophysics,
Flatiron Institute, 162 5th Avenue,
New York, NY 10010, USA}

\begin{abstract}

Turbulent radiative mixing layers (TRMLs) play an important role in many astrophysical contexts where cool (${\la}10^4$  K) clouds interact with hot flows (e.g., galactic winds, high velocity clouds, infalling satellites in halos and clusters). 
The fate of these clouds (as well as many of their observable properties) is dictated by the competition between turbulence and radiative cooling; however, turbulence in these multiphase flows remains poorly understood.
We have investigated the emergent turbulence arising in the interaction between clouds and supersonic winds in hydrodynamic \enzoe\ simulations. 
In order to obtain robust results, we employed multiple metrics to characterize the turbulent velocity, \vturb.
We find four primary results, when cooling is sufficient for cloud survival.
First, \vturb\ manifests clear temperature dependence.
Initially, \vturb\ roughly matches the scaling of sound speed on temperature.
In gas hotter than the temperature where cooling peaks, this dependence weakens with time until \vturb\ is constant.
Second, the relative velocity between the cloud and wind initially drives rapid growth of \vturb.
As it drops (from entrainment), \vturb\ starts to decay before it stabilizes at roughly half its maximum.
At late times cooling flows appear to support turbulence.
Third, the magnitude of \vturb\ scales with the ratio between the hot phase sound crossing time and the minimum cooling time.
Finally, we find tentative evidence for a length-scale associated with resolving turbulence.
Under-resolving this scale may cause violent shattering and affect the cloud's large-scale morphological properties.
\end{abstract}


\section{Introduction} \label{sec:intro}

While scales and relevant physics may vary, interactions between regions of cooler gas and coherent flows of hotter gas are prominent in many contexts.
These interactions are prevalent in the circumgalactic medium (CGM), such as high velocity clouds \citep[e.g.,][]{Wakker:1997, Putman:2012}, ram-pressure stripping of infalling satellites \citep[e.g.,][]{emerick2016, simons20a} and the resulting streams \citep[e.g.,][]{bland-hawthorn2007,  bustard22a}, or cooling flows from cosmic accretion \citep[e.g.][]{mandelker20a}.
There are also instances of these interactions within the interstellar medium (ISM), like the stellar-wind driven bubbles within star-forming clouds \citep[e.g.][]{lancaster21a}.
They are also relevant to the ram-pressure stripping of cluster galaxies and star formation in the tails of jellyfish galaxies \citep[e.g.][]{tonnesen21a}.
We take a particular interest in their role within galactic winds \citep[e.g.][]{fielding22a}.

Galactic winds are ubiquitous throughout cosmic time, and play a pivotal role in galaxy evolution; they regulate star formation and transport metals out of the interstellar medium (ISM) \citep{somerville15a}.
Observations indicate that stellar-feedback-driven winds are inherently multiphase; they are composed of comoving gas phases that vary in temperatures by orders of magnitude (see \citealp{veilleux05a} and \citealp{rupke18a} for reviews of observational evidence). 

Observations favor a model in which supernovae drive hot ${\ga}10^6\, {\rm K}$ winds that accelerate and entrain clouds of cool ${\sim}10^4\, {\rm K}$ gas from the ISM \citep[e.g.][]{chevalier85a}.
This model is complicated by hydrodynamical instabilities that drive mixing of gas between the cloud and wind.
Because the timescale for mixing to destroy the cloud (by homogenizing the gas phases) is shorter than the ram-pressure acceleration timescale, it's remarkably difficult to accelerate clouds before they're destroyed \citep{zhang17a}.

Various ideas have been proposed to address this difficulty.
Some, like magnetic shielding \citep[e.g.][]{mccourt15a,gronnow18a,cottle20a}, may extend the cold-phase lifetime by reducing mixing \citep[see also][for other mechanisms]{forbes19a}.
Others are alternative acceleration mechanisms like radiation-pressure \citep[e.g.][]{zhang18a} or cosmic rays \citep[e.g.][]{wiener19a,bruggen20a}.
Another idea suggests the remnants of destroyed clouds seed the in situ formation of clouds in cooling outflows \citep[e.g.][]{thompson15a, schneider18a, lochhaas21a}.

Radiative cooling is also known to extend the cold-phase lifetime \cite[e.g.][]{mellema02a, fragile04a, melioli05a,cooper09a}. This work focuses on the regime in which rapid cooling acts as a mechanism that facilitates cloud survival \citep[e.g.][]{marinacci10a,armillotta16a}.
In this regime, cooling in a thin layer of gas at the interface between the phases is able to overcome the destructive effects of mixing \citep{gronke18a}.
As turbulent mixing feeds hot phase material into this layer, isobaric cooling removes the temperature differential in the new material \citep{fielding20a}.
This process facilitates the transfer of mass and momentum to the cold phase providing a powerful additional acceleration source and allowing cloud growth.
Hereafter, we refer to this mechanism as turbulent radiative mixing layer (TRML) entrainment.

This topic has been extensively studied using wind tunnel setups \citep[e.g.][]{gronke20a, li20a, sparre20a, kanjilal20a, abruzzo22a, bustard22a, farber22a}.
There has also been considerable work that focuses on a single shear layer \citep[e.g.][]{Kwak:2010,ji19a,fielding20a,tan21a}.

The literature largely agrees that the occurrence and efficacy of TRML entrainment is controlled by three \pdns: (i) the density contrast $\chi=\rhocl/\rhow$ between the cloud and the wind, (ii) the Mach number of the wind $\Mw=\vw/\cshot$, and (iii) the cooling efficiency $\xi=\tau_{\rm mix}/\tau_{\rm cool}$.
Here, $\tau_{\rm mix}$ and $\tau_{\rm cool}$ specify the characteristic timescales for mixing and for cooling of the mixing layer.
As in \citet{fielding20a} we primarily consider $\xish=\tshear/\tcoolmin$, where $\tshear=\rcl/\vw$ is the shear-time and \tcoolmin\ is the minimum cooling time.
In practice, our choice for $\tau_{\rm cool}$ is similar to the popular option of using \tcoolmix, the cooling time of gas within the mixing layer at $\Tmix\sim\sqrt{\Tcl\Tw}$ and $\nmix\sim\sqrt{\ncl\nw}$.\footnote{As in \citet{abruzzo22a}, we actually define $\emix=\sqrt{\ecl\ew}$, where $e$ is the specific internal energy.
This definition is more consistent with the arguments of \citet{begelman90a}, when the mean molecular weight is not constant.
Since the problem is quasi-isobaric, $\rhomix=\sqrt{\rhocl\rhow}$.
Consequently, the geometric mean of \Tcl\ and \Tw\ (\ncl\ and \nw) tends to slightly overestimate (underestimate) the value of \Tmix\ (\nmix).}
It has been suggested that the relevant cooling timescale is instead set by cooling in the hot, volume filling, wind phase \citep{li20a,sparre20a}. 
We reconcile differences between these cooling timescales in follow-up work \citepAbruzzoSurvival.

Despite the obvious central importance of turbulence mechanisms underlying the operation of TRMLs, we do not yet have a clear understanding of how the turbulent velocity $\vturb$ changes as the 3 \pdns\ ($\chi$, $\Mw$, and $\xi$) are varied. This is closely related to two fundamental unanswered questions. 

(i) \emph{What is the role of cooling in driving turbulence?} Shear layer studies find no or very weak cooling time dependence of $\vturb$ \citep{fielding20a,tan21a}. In contrast, some cloud crushing simulations find that cooling induced pulsations may be the dominant driver of turbulence \citep{gronke20a,gronke20b}. Reconciling these pictures requires a careful investigation of how $\vturb$ scales with $\xi$.

(ii) \emph{What is the timescale for turbulent mixing?} Shear layer studies associate $\tau_{\rm mix}$ with the eddy turnover time at the outer scale, or $\tau_{\rm mix}\sim L_{\rm outer}/\vturb(\Tcl,\ell=L_{\rm outer})$, where $\vturb(\Tcl,\ell=L_{\rm outer})$ is a fixed fraction of the relative velocity for $\chi\ga100$ \citep{fielding20a,tan21a}.
This scales similarly to \tshear.
Wind-tunnel studies instead link $\tau_{\rm mix}$ with the cloud-crushing time, $ \tcc=\sqrt{\chi}\rcl/\vw$.
The Kelvin-Helmholtz and Rayleigh-Taylor instabilities have growth times of order \tcc\ and destroy clouds over a few \tcc, in the absence of cooling \citep{klein94a}.
\citet{gronke18a} predicts cloud survival when $\xigo=\tcc / \tcoolmix$ exceeds unity.
While both choices give $\xi$ a $\vw \rcl^{-1}$ scaling, the latter introduces an extra dependence on $\chi^{-1/2}$. This discrepancy could have profound impacts on cloud survival criteria and requires a careful understanding of how $\vturb$ scales with $\chi$.

To address these questions, we investigate the turbulent properties that emerge in wind-tunnel simulations of cloud-wind interactions.
While turbulence in TRMLs has traditionally been treated as homogeneous \citep[e.g][]{begelman90a,gronke18a,fielding20a}, we will show that it depends not just on scale but also on phase.
This has important implications for mixing and hence cloud survival.
Although most previous work on TRML entrainment has focused on cloud-wind density contrasts of $\chi = 100-300$ \citep[see however][]{sparre20a,gronke18a,gronke20a}, galactic winds are expected to have $\chi \gtrsim 10^4$ \citep{fielding22a}. Furthermore, we have preliminary evidence for important changes to the dynamics and clumping structure for $\chi \gg 10^2$ \citep{gronke20b}.
In this work we, therefore, place particular emphasis on higher $\chi$ results.

In \autoref{sec:simulations}, we describe the suite of simulations used in this investigation.
Videos of these simulations can be found at \url{http://matthewabruzzo.com/visualizations/}.
In \autoref{sec:characterize_turb}, we describe and compare three approaches for characterizing multiphase turbulence, followed by a description of the results from applying these methods to our simulation suite in \autoref{sec:results}.
Subsequently, we describe implications of our results and detail our conclusions in \autoref{sec:discussion} and \autoref{sec:conclusion}.

\begin{figure*}[ht]
  \center
\includegraphics[width =\textwidth]{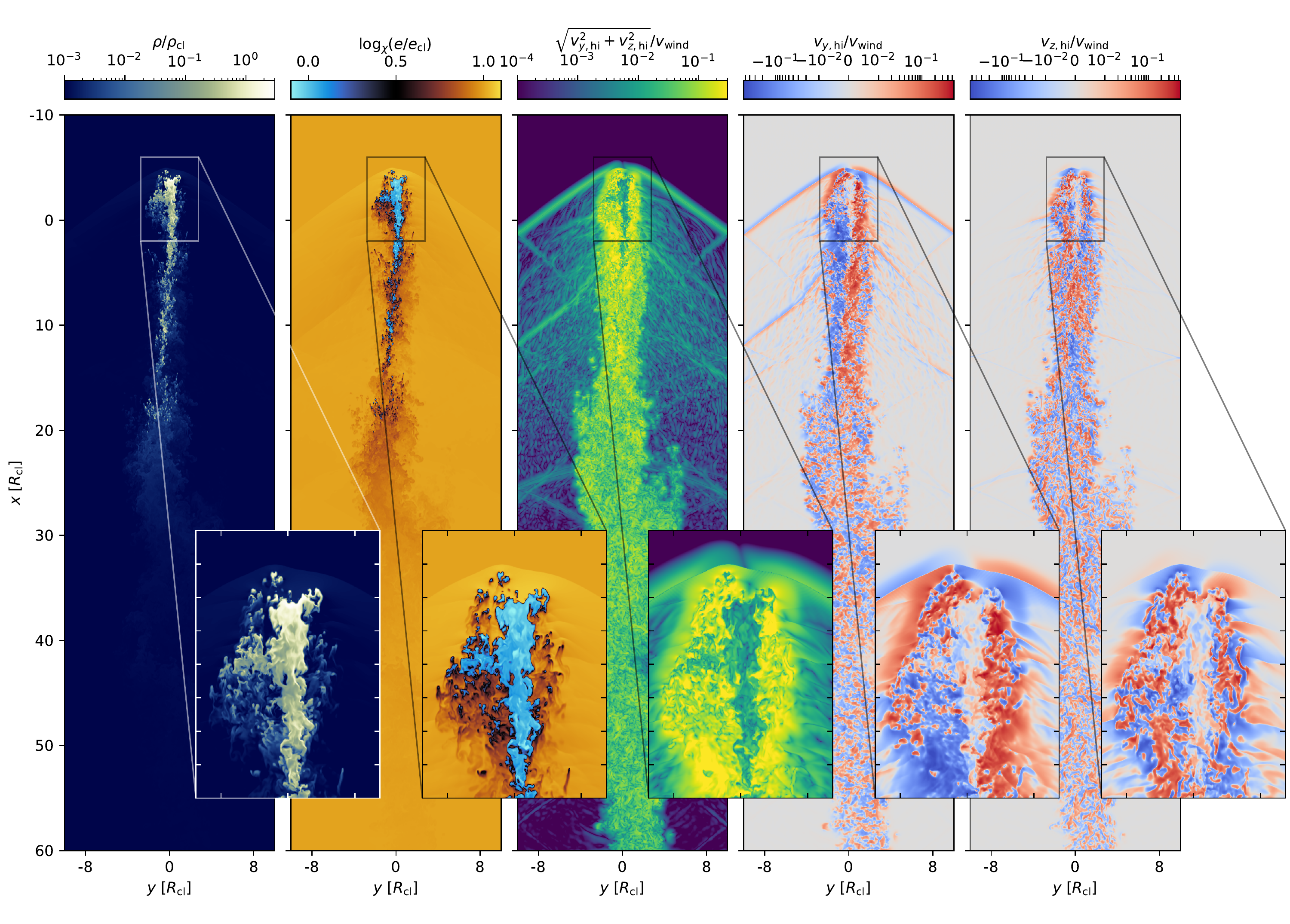}
\vspace{-0.3in}
\caption{\label{fig:X1000slice}
  Slice of a $\chi = 1000$, $\xish = 27.8$, $\Mw=1.5$ simulation at $4.5\tcc$.
  This simulation has a resolution of 64 cells per cloud radius and the cloud is eventually entrained in the wind.
  The left two panels show the density and specific internal energy, which is $T/\mu$ scaled by physical constants.
  The right two panels show the high-pass filtered components of the velocity field transverse to \vwindhat\ and the center panel shows the combined magnitude of these values.
  The insets highlight how the turbulent velocity has a clear temperature dependence.
}
\end{figure*}

\begin{deluxetable*}{ccccccccl}[ht!]
\tabletypesize{\footnotesize} %
\tablecaption{\label{tab:sim_table} Table of simulations.}
\tablehead{
  \colhead{$\chi$} &
  \colhead{$\mathcal{M}_w$} &
  \colhead{$R_{\rm cl}$ (pc)} &
  \colhead{$t_{\rm cc}/t_{\rm cool,mix}$} &
  \colhead{$t_{\rm shear}/t_{\rm cool,min}$} &
  \colhead{Survival?\tablenotemark{a}} &
  \colhead{$R_{\rm cl} / \Delta x$} &
  \colhead{Domain ($R_{\rm cl}$)} &
  \colhead{notes}
}
\startdata
100 & 1.5 & 8.647 &  &  & No & 16 & $120\times 10^2$ & No cooling \\
100 & 1.5 & 5.638 & 1.34 & 0.57 & Borderline\tablenotemark{b} & 8,16 & $120\times 10^2$ &  \\
100 & 1.5 & 12.1 & 2.87 & 1.23 & Yes & 16 & $120\times 10^2$ &  \\
100 & 1.5 & 22.4 & 5.32 & 2.28 & Yes & 16 & $120\times 10^2$ &  \\
100 & 1.5 & 26.75 & 6.35 & 2.72 & Yes & 16 & $120\times 10^2$ &  \\
100 & 1.5 & 44.12 & 10.48 & 4.49 & Yes & 16 & $120\times 10^2$ &  \\
100 & 1.5 & 56.38 & 13.39 & 5.73 & Yes & 4,8,16,32 & $120\times 10^2$ &  \\
100 & 1.5 & 121.0 & 28.73 & 12.30 & Yes & 16 & $120\times 10^2$ &  \\
100 & 1.5 & 262.0 & 62.21 & 26.64 & Yes & 16 & $120\times 10^2$ &  \\
100 & 1.5 & 441.2 & 104.76 & 44.86 & Yes & 16 & $120\times 10^2$ &  \\
100 & 1.5 & 563.8 & 133.87 & 57.33 & Yes & 16 & $120\times 10^2$ &  \\[1.5mm]
100 & 0.75 & 28.19 & 13.39 & 5.73 & Yes & 16 & $120\times 10^2$ &  \\
100 & 3.0 & 112.76 & 13.39 & 5.73 & Yes & 16 & $240\times 10^2$ &  \\
100 & 3.0 & 1127.6 & 133.9 & 57.3 & Yes & 16 & $240\times 10^2$ &  \\
100 & 6.0 & 225.52 & 13.39 & 5.73 & Yes & 16 & $240\times 10^2$ &  \\[1.5mm]
100 & 1.5 & 77.27 & 10.00 & 5.55 & Yes & 16 & $120\times 10^2$ & $T_{\rm cl} = 8910\, {\rm K}$ \\
100 & 1.5 & 121.0 & 15.66 & 8.70 & Yes & 16 & $120\times 10^2$ & $T_{\rm cl} = 8910\, {\rm K}$ \\
100 & 1.5 & 316.58 & 10.00 & 15.89 & Yes & 16 & $120\times 10^2$ & $T_{\rm cl} = 15845\, {\rm K}$ \\[1.5mm]
300 & 1.5 & 13.67 & 2.50 & 0.80 & No & 16 & $120\times 10^2$ &  \\
300 & 1.5 & 27.34 & 5.00 & 1.61 & No & 16 & $120\times 10^2$ &  \\
300 & 1.5 & 54.68 & 10.00 & 3.21 & Yes & 16 & $120\times 10^2$ &  \\
300 & 1.5 & 164.0 & 29.99 & 9.63 & Yes & 16 & $120\times 10^2$ &  \\
300 & 1.5 & 546.8 & 100.01 & 32.10 & Yes & 16 & $120\times 10^2$ &  \\[1.5mm]
$10^3$ & 1.5 & 8.647 &  &  & No & 16 & $120\times 10^2$ & No cooling \\
$10^3$ & 1.5 & 86.47 & 10.00 & 2.78 & No & 16 & $240\times 10^2$ &  \\
$10^3$ & 1.5 & 173.0 & 20.01 & 5.56 & Borderline & 16 & $240\times 10^2$ &  \\
$10^3$ & 1.5 & 324.0 & 37.47 & 10.42 & Borderline\tablenotemark{c} & 16 & $240\times 10^2$ &  \\
$10^3$ & 1.5 & 432.35 & 50.00 & 13.90 & Yes & 16 & $240\times 10^2$ &  \\
$10^3$ & 1.5 & 864.7 & 100.00 & 27.81 & Yes & 4,8,16,32,64 & $240\times 10^2$\tablenotemark{d} &  \\
$10^3$ & 1.5 & 2594.1 & 300.01 & 83.42 & Yes & 16 & $240\times 10^2$ &  \\[1.5mm]
$10^4$ & 1.5 & 8.647 &  &  & No & 16 & $360\times 30^2$ & No cooling \\
$10^4$ & 1.5 & 1699.5 & 10.00 & 17.28 & No & 16 & $360\times 30^2$ & Wind can cool \\
$10^4$ & 1.5 & 16995.0 & 100.00 & 172.82 & Borderline\tablenotemark{e} & 8,16 & $360\times 30^2$ & Wind can cool \\
\enddata
\tablecomments{
    Unless otherwise noted, all runs were initialized with $\Tcl=5010\, {\rm K}$, where $\tcoolcl/\tcoolmin\sim105$.
    All simulations were initialized with an initial thermal pressure of $p/k_B=10^3\, {\rm cm}^{-3}\, {\rm K}$.
    In each run, \tcool\ is minimized at $T=1.83 \times 10^4\, {\rm K}$ with a value of $75.5\, {\rm kyr}$; the sound speed at this temperature is 18.6 km/s.
    The cooling length, $\lcool=\cs\tcool$, is minimized at $T=1.70 \times 10^4\, {\rm K}$ with a value of $1.43\, {\rm pc}$.
}
\tablenotetext{a}{Denotes whether clouds survive (i.e. if the cold phase mass ever drops to 0). ``Borderline'' indicates cases where the line between survival vs. destruction and rapid subsequent precipitation is fuzzy}
\tablenotetext{b}{the cold phase mass, $\rho>\sqrt{\chi}\rhocl$, dropped to ${\sim}0.01\%,\ 0.07\%$ of the initial value in the $R_{\rm cl} / \Delta x=8,16$ runs before growth. At these times, there is no mass denser than $\rhocl/3$.}
\tablenotetext{c}{the mass of gas denser than $\sqrt{\chi}\rhocl$ ($\rhocl/3$) drops to $16\%$ ($6\%$) of it's original value and begins monotonic growth after $21.5{\tcc}$ ($24{\tcc}$).
In an alternate version of the same run, where the domain dimensions are $120\times 10^2$, the mass instead drops to $32\%$ ($8\%$) of its initial value and starts growing after $12.5{\tcc}$ ($11{\tcc}$).
}
\tablenotetext{d}{$R_{\rm cl} / \Delta x\geq32$ runs used domains with dimensions $120\times 10^2$}
\tablenotetext{e}{the mass of gas denser than $\sqrt{\chi}\rhocl$ ($\rhocl/3$) drops to ${\sim}0.01\%, 26\%$ ($0\%, 4\%$) of the initial value in the $R_{\rm cl} / \Delta x=8,16$ runs. 
Only the higher resolution case shows significant subsequent growth.}
\end{deluxetable*}

\section{Simulations} \label{sec:simulations}

We ran a suite of 3D uniform grid hydrodynamical simulations using the \enzoe\footnote{\url{http://enzo-e.readthedocs.io}} code,
which is a rewrite of \enzo\ \citep{bryan14a} built on the adaptive mesh
refinement framework \cello\ \citep{bordner12a,bordner18a}.
Our simulations employed the van Leer integrator (without constrained transport) \citep{stone09a} with second order reconstruction and the HLLC Riemann solver.

Our simulations begin with a motionless spherical cloud embedded within a hot, uniform, laminar wind in the $\hat{x}$ direction.
We imposed an inflow condition on the upstream boundary (positive $\hat{x}$) and outflow conditions for the other boundaries.
The cloud and wind material are initialized with $p/k_B = 10^3\, {\rm K}\, {\rm cm}^{-3}$.
The cloud density \rhocl\ in all of our simulations is chosen such that $\Tcl = 5010\, {\rm K}$. This roughly corresponds to the temperature where heating starts to dominate over cooling (without self-shielding). The wind density is then determined by the desired value of $\chi$.

To model radiative cooling, we use the \grackle\footnote{\url{https://grackle.readthedocs.io/}} library \citep{smith17a}, assuming solar metallicity and no self-shielding.
Specifically, we use the tabulated heating and cooling rates for optically thin gas in ionization equilibrium with the $z = 0$ \citet{haardt12a} UV background.
We turn off cooling in gas with $T>0.6 \Tw$ in our simulations with $\chi \leq 10^3$.
This helps to avoid complications from cooling in the hot wind in our $\chi = 100$ simulations in which the ratio of the cooling time of the hot wind to the cooling time of the mixing layer is $\tcoolw/\tcoolmix\sim40$.
In higher $\chi$ simulations cooling of the wind fluid is so slow that this ceiling has no discernible impact.\footnote{
We only explicitly checked the effects of a cooling wind in the $\chi=100,\Mw=1.5$ runs and $\chi=10^3,\xish=27.8$ run.
We expect only minimal late time complications in our $\chi=1000,\xish=83.4$ run since it has $\tcoolw/\tcc\sim51$ \citep{abruzzo22a}.
While our $\chi=300,\xish=3.2$ run has the same $\tcoolw/\tcc$, complications may be significant since that run is close to the survival threshold.
Complications are likely significant in larger $\chi=300$ runs.
}
For simplicity, we also turned off heating/cooling below \Tcl.

To break initial symmetries, we initialized the density of each cell, within the cloud, to the average of $\rho (\Vec{x})$, where
\begin{equation}
    \frac{\rho (\Vec{x})}{\rho_{\rm cl}} = 1 + 0.099 \sum_{i=1}^{10} \cos \left(\frac{2 \pi}{\lambda_{i}} \hat{e}_i \cdot \Vec{x} + \phi_i\right).
\end{equation}
For each $i$, we drew a random unit vector $\hat{e}_i$ and values for $\lambda_i$ and $\phi_i$ from $[R_{\rm cl}/8,R_{\rm cl}]$ and $[0,\pi)$.
Cells on the cloud edges were initialized with subsampling; each subcell had a width of $\rcl/128$.

Our simulations have resolutions of $\rcl/\cellwidth =4,8,16,32,64$.
Unless stated otherwise, results are presented for $\rcl/\cellwidth = 16$.
By default, the wind-aligned dimension and transverse dimensions for most of our simulations' domains had sizes of $120{\rcl}$ and $20{\rcl}$, respectively corresponding to a $1920 \times 320^2$ grid at our fiducial $\rcl/\cellwidth = 16$ resolution.
The sizes were somewhat larger ($360{\rcl}$ and $30{\rcl}$) for our $\chi=10^4$ simulations in order to minimize the impact of the bow shock reflections, prevent dense material from leaking out of the transverse boundaries in cases of shattering, and to give room for tail formation.
While the default dimensions are adequate for determining whether our clouds survive in runs with $\Mw \geq3$ or $\chi=10^3$, we find that boundary effects can impact later time measurements.
Thus for such cases, with radiative cooling and $\rcl/\cellwidth = 4,8,16$, we present results from runs with a wind-aligned length of $240{\rcl}$.
In all cases, the cloud was initialized at the center of the domain and we employed a frame-tracking scheme that updated the reference frame every $\tcc/16$ such that the mass-weighted velocity for cells with $\rho \geq \sqrt{\rhocl \rhow}$ was zero.

\autoref{tab:sim_table} presents a list of our simulations.

As we will discuss in \autoref{sec:characterize_turb}, our measurements of \vturb\ involve averages over velocity properties.
Thus, leakage of material from the domain could plausibly bias our measurements.
However, the generality of the scaling relations derived in this work, which apply to runs that do and do not leak material, suggests that overall effects on our \vturb\ measurements are probably minimal. 
We assessed this leakage by tracing material initialized within the cloud with a passive scalar.
Nearly all turbulent measurements shown in this work for our $\rcl/\cellwidth =16$ runs, where the cloud avoids destruction, are from times at which our runs retain at least $95\%$ of the passively advected scalar.
This statement doesn't apply to our $\chi=300,\xish=3.21$ ($\chi=10^3,\xish=10.47$; $\chi=10^3,\xish=13.9$) run, which retains $95\%$ of the scalar until $11{\tcc}$ ($11.5{\tcc}$; $12{\tcc}$), and leaks another ${\sim}6\%$ (${\sim}34\%$; ${\sim}17\%$) by $14.5{\tcc}$ ($17.5{\tcc}$; $18.5{\tcc}$).
Additionally, our $\chi=10^3,\xish=27.81$ run retains $95\%$ for $22{\tcc}$, but only loses another ${\sim}2\%$ over the subsequent $18{\tcc}$. 
Finally, our $\chi=10^4,\xish=178.82$ case retains $95\%$ for $8{\tcc}$, but only retains $42\%$ ($15\%$) by $15{\tcc}$ ($27.5{\tcc}$).
This last case is particularly noteworthy because it starts leaking the scalar at $6.5{\tcc}$, which coincides with a drop in the cool phase mass.

\section{Characterizing Turbulence} \label{sec:characterize_turb}

The primary goal of this paper is to characterize the turbulent properties of the turbulent radiative mixing layer that mediates mixing and cooling between the hot wind and cold cloud. Although much effort has been devoted to understanding turbulence in single phase media, there has been considerably less work for multiphase systems \citep[e.g.,][]{mohapatra22_filaments,gronke22a,gronke22b}.
The potential dependence of turbulent properties on both scale and the gas' local thermodynamic state complicates the interpretation of conventional methods for characterizing \vturb. There are a number of possible ways to extend existing turbulence measures; however, their novel nature means that they can be difficult to interpret and their robustness is unclear. In order to get around this difficulty, in this paper we consider three distinct methods for characterizing our multiphase turbulent simulations. These are built around three different ideas based on (i) a filter-based technique, (ii) a geometric approach, and (iii) classic structure function ideas. 

We describe these approaches below, and to supplement our description of the methods, we apply each to a snapshot of a $\rcl/\cellwidth=64$ of $\chi = 1000$, $\xish = 27.8$, $\Mw=1.5$, which simulates a successfully entrained cloud.

\subsection{Filtering}
\label{sec:hipass}

\begin{figure}
  \center
\includegraphics[width = 3.35in]{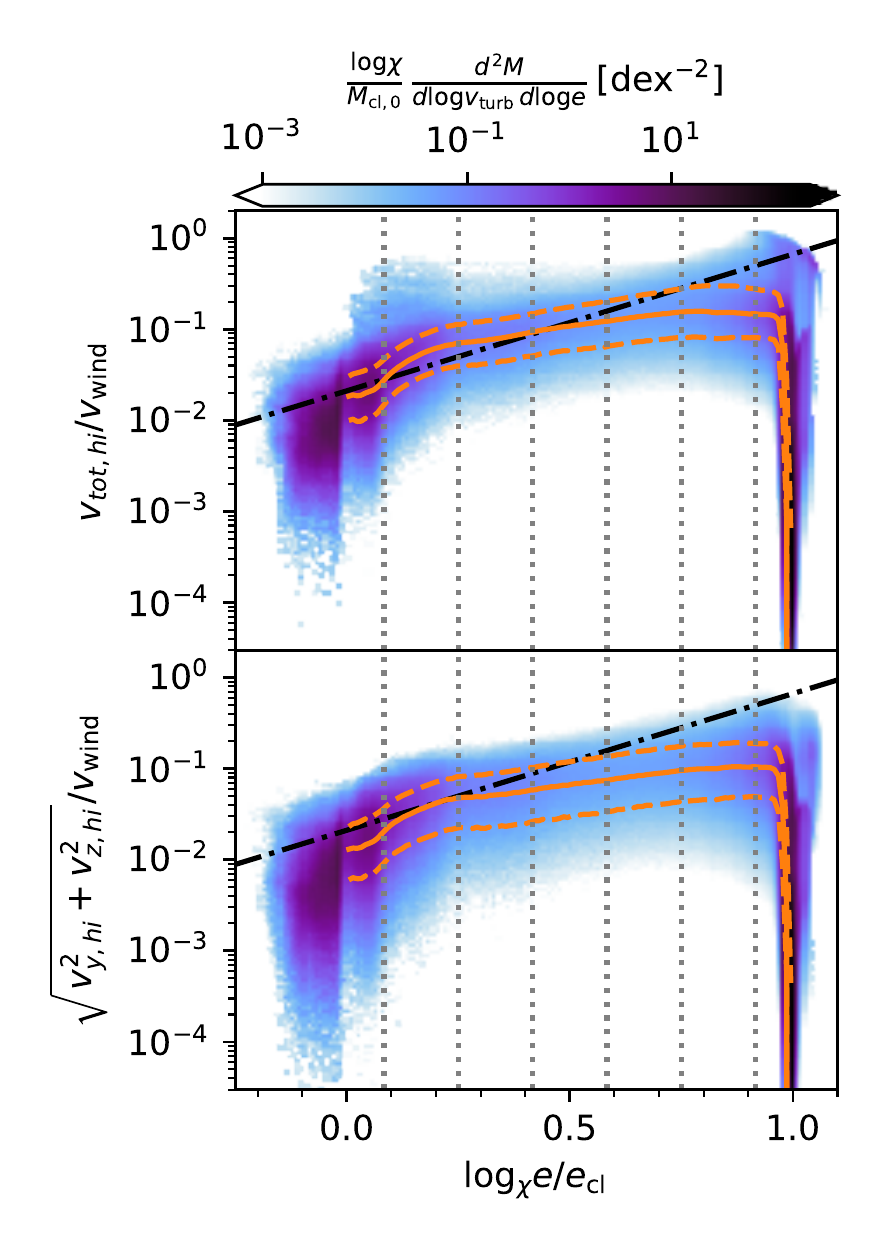}
\caption{\label{fig:filter_phase} Shows the phase dependence of \vturb, measured via filtering, of the $\rcl=64\cellwidth$ run of our $\chi = 1000$, $\xish = 27.8$, $\Mw=1.5$ simulation at  \fidtime.
The top panel includes contributions from all three velocity components.
The bottom panel just includes contributions from the components transverse to \vwindhat; this is consistent with how \vturb\ from filtering is measured throughout the remainder of this work.
The solid orange line denotes the median while the dashed orange lines bound values between the 15th and 85th percentile.
The dotted-dashed line shows \vturb\ magnitudes that are equal to the sound-speed.
The steep drop-off in \vturb\ near \Tw, is an artifact of the fact that wind is initially laminar.
}
\end{figure}

In our first approach, we attempt to explicitly remove the bulk flows by filtering out the large-scale bulk velocities.
Specifically, we estimate \vturb\ by applying a high-pass Gaussian filter, with density weighting, to each component of the velocity. The size of the filter is chosen to correspond to scales on which the bulk flow is varying, that is approximately the cloud radius.
We use density weighting (which corresponds to smoothing the momentum) in order not to be dominated by the volume-filling hot gas component.

More precisely, in this approach, the $i$-th component of the turbulent velocity is given by
\begin{equation}
    \label{eqn:hipass}
    v_{i,{\rm turb}}(\Vec{x}) = v_i(\Vec{x}) - 
    \frac{\iiint f_\sigma(\Vec{x} - \Vec{r}) \rho(\Vec{r}) v_i(\Vec{r}) d^3\Vec{r}}
    {\iiint f_\sigma(\Vec{x} - \Vec{r}) \rho(\Vec{r}) d^3\Vec{r}},
\end{equation}
where $f_\sigma(\Vec{x})$ is the formula for a normalized, separable, three-dimensional Gaussian.
In short, the convolution of $f_\sigma(\Vec{x})$ with $v_i(\Vec{r})$ (i.e. the fraction term) estimates the laminar part of $v_i$, and subtracting if from $v_i$ gives the turbulent part.

Throughout this work, we use a Gaussian filter with a standard deviation of $\rcl /4$; this was chosen after extensive experimentation to visually pick out turbulent regions with a minimal ``bleed" into the laminar regions.
Our results do not depend qualitatively on the exact choice of the filtering scale as long as it is on the order of the cloud size.

The rightmost two panels in \autoref{fig:X1000slice}, illustrate the hi-pass filtered transverse velocity components for the aforementioned simulation at $4.5{\tcc}$, and the center panel shows the combined magnitude of these components, $v_{\rm trans,hi}$.
The left two panels show the density and specific thermal energy slices.
Note that here, as elsewhere in this paper, we use $e \equiv (p/\rho)/(\gamma-1)$ to denote the specific thermal energy of the gas.
This quantity is closely related to temperature, but is easier to compare among runs with different $\chi$ values since $e_{\rm wind}=\chi e_{\rm cl}$ (due to variations in mean molecular mass $\Tw < \chi \Tcl$).
The inset panels make it readily apparent that the turbulent velocity has a clear phase dependence.

In \autoref{fig:filter_phase} we show the phase dependence explicitly (at $2.5{\tcc}$), plotting the 2D distribution of mass as a function of temperature and (top) hi-pass filtered velocity including all components, $v_{\rm tot,hi}$, and (bottom) just the transverse components, $v_{\rm trans,hi}$ for this snapshot.
Because of the spatial gradients that persist in the downstream velocity component (see \autoref{appendix:early-time-vturb}), we use $v_{\rm trans,hi}$ to estimate \vturb\ in the remainder of this work.

This approach is sensitive to turbulence on scales below the high-pass filtering limit; since this is approximately the driving scale (of order the cloud radius), we expect this to be a good measure of the turbulent properties, although it may also remove some of the contribution to the turbulence on scales just below the driving scale.
One possible downside of this approach is the contribution of the bulk flow in scales at and below the filtering scales; we have explored alternate weighting schemes and find only minor differences.
Although we do not have detailed scale information (except the removal of large scales), this approach does permit a very fine examination of the turbulent properties with phase (i.e., specific energy) as seen in \autoref{fig:filter_phase}.

Indeed, this figure clearly shows a different dependence on specific energy below and above $\log_{10}(e/e_{\rm cl}) \approx 0.7$, which corresponds to the peak of the cooling curve we adopt.
We return to this point in \autoref{sec:results}.
\autoref{fig:X1000slice} qualitatively shows that spatial variations in turbulence are largely explained by the spatial variations in gas phase.
The main exception is the hottest phase, which is ``contaminated'' by unmixed, laminar gas (this is reflected in \autoref{fig:filter_phase}).

Because measuring phase information isn't as seamless for our other approaches, we define a set of nominal coarse phase bins to be used with them.
We define the bin edges in terms of $\log(e/e_{\rm cl}) / \log\chi$ to ease comparisons between runs with different $\chi$ values.
The bin edges are $-\infty$, $1/12$, $3/12$, $5/12$, $7/12$, $9/12$, $11/12$, which are illustrated by the vertical dotted lines in \autoref{fig:filter_phase}.

\subsection{Geometric}

\begin{figure*}
\center
\includegraphics[width = 6.5in]{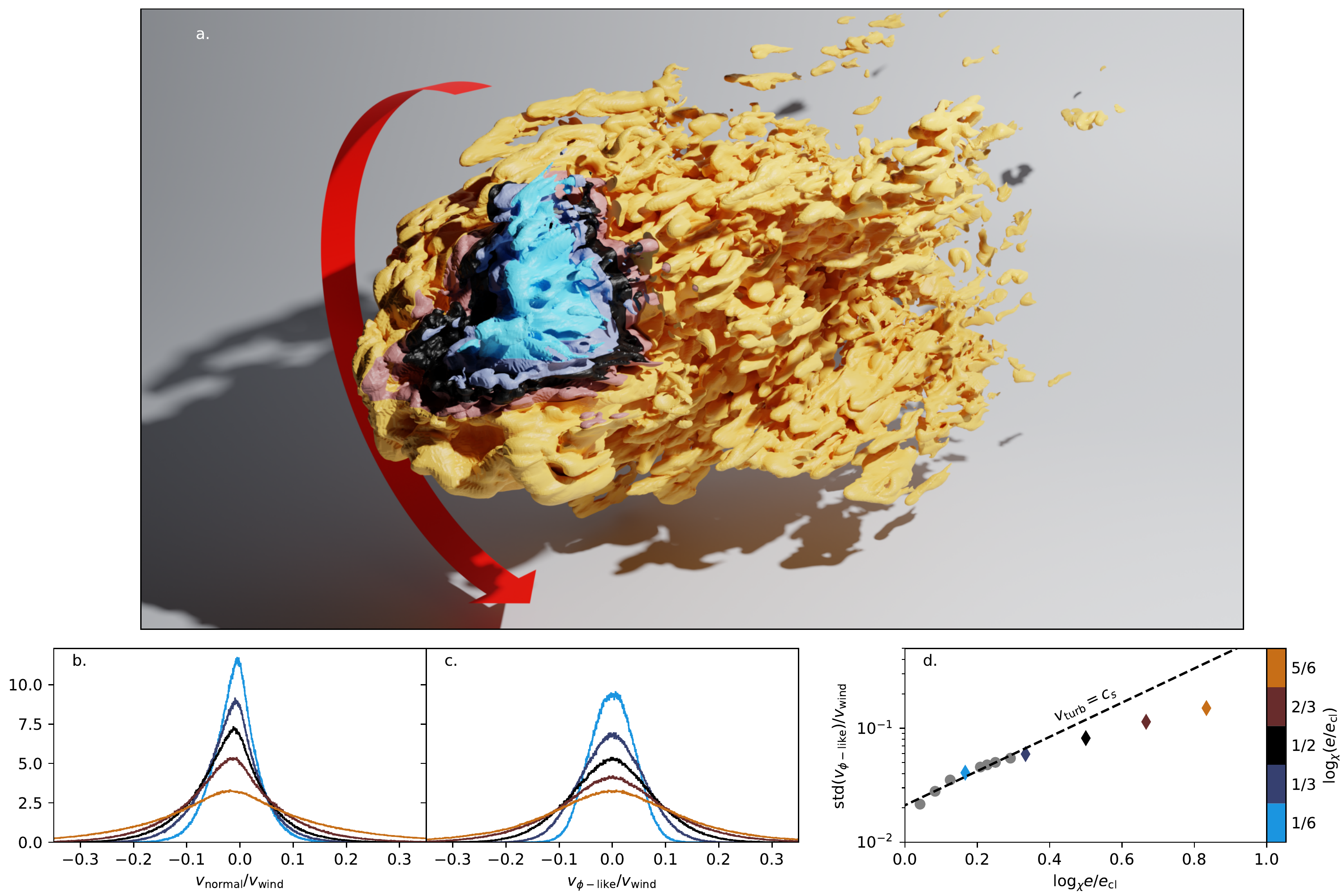}
\caption{\label{fig:isosurface} Illustrates iso-temperature surfaces and derived \vturb\ measurements for our the $\rcl/\cellwidth=64$ run of our $\chi = 1000$, $\xish = 27.8$, $\Mw=1.5$ simulation at $2.5\tcc$. 
Panel~a shows a cut-away of five nested iso-surfaces measured at $\logXe={1/6, 1/3, 1/2, 2/3, 5/6}$ (for this system, $T ={1.3\times10^4\, {\rm K},\ 3.3\times10^4\, {\rm K},\ 10^5\, {\rm K},\ 3.3\times10^5\, {\rm K},\ 3.3\times10^4\, {\rm K},\
10^6\, {\rm K}}$).
The arrow illustrates the $\hat{\phi}$ direction measured in the plane transverse $\Vec{\vw}$.
Panels b and c respectively show the normalized area-weighted distributions of the $v_{\rm normal}$ and \vphilike\ velocity components measured on the iso-surfaces pictured in a.
Panel d shows the standard deviation of the distributions from panel c (colored diamonds), as well as data derived from other isosurfaces (gray circles), plotted as a function of \logXe.
} 
\end{figure*}

Our second approach uses the geometry of isosurfaces in the flow to characterize the turbulence.
To motivate this, consider a toy model in which a cloud's geometry is a sphere or a cylinder. 
The cloud is oriented such that the azimuthal angle, $\phi$, measures the angle on the plane transverse to $\Vec{v_{\rm wind}}$.
While cloud acceleration and accretion \citep[e.g. by a TRML,][]{fielding20a} can drive steady coherent flows along the wind and radial directions, turbulence is the only source of motion along $\hat{\phi}$.
In other words, we can characterize \vturb\ with $v_\phi$.
\citet{tan21a} drew a similar conclusion in shearing box simulations about the utility of the dispersion of the velocity component perpendicular to shear and inflow directions.

Despite their more complex morphology, we can apply the same logic to real clouds.
For a given snapshot, we employ the \citet{lewiner03a} marching cubes algorithm to construct five topologically correct meshes of triangle facets that trace specific internal energy isosurfaces using values that coincide with the centers of the closed bins mentioned in \autoref{sec:hipass}.
We supplement these with additional isosurfaces at values near the peak of the cooling curve (we vary the precise locations based on the $\chi$ value of the simulation). 
\autoref{fig:isosurface}a shows a cutaway visualization of several of these iso-surfaces at \fidtime\ for our $\chi = 1000$, $\xish = 27.8$ simulation.

For each facet, we define $\vphilike  \equiv \Vec{v} \cdot \left(\vwindhat \times \hat{n}\right)$, where \Vec{v} is the linearly interpolated velocity and $\hat{n}$ is the outward normal vector.
Finally, we estimate \vturb\ for an iso-surface with the area-weighted standard deviation of \vphilike (excluding facets with $\vwindhat \times \hat{n}=\Vec{0}$).

\autoref{fig:isosurface}b-c shows area-weighted distributions of $\vrlike \equiv \Vec{v}\cdot\hat{n}$ and \vphilike\ for the previously mentioned simulation.
The distribution for \vrlike\ shows a negative mean for each iso-surface, which is consistent with net-inflow of gas.
In contrast, the distributions for \vphilike\ is centered on 0, which is exactly what we expect.

While this approach does not provide any scale information about turbulence, it can be used to provide detailed phase information.
For example, \autoref{fig:isosurface}d illustrates qualitatively similar phase-dependence to the filtering measurements.
However, in contrast to the filtering technique, this approach requires generation of a separate surface for each phase to be probed and so is much more computationally intensive.
As is discussed in \autoref{appendix:early-time-vturb}, the main advantage of this approach is that it provides the most accurate early-time measurements.

\subsection{Velocity Structure Function} \label{sec:vsf}

\begin{figure}
  \center %
  \includegraphics[width = 3.35in]{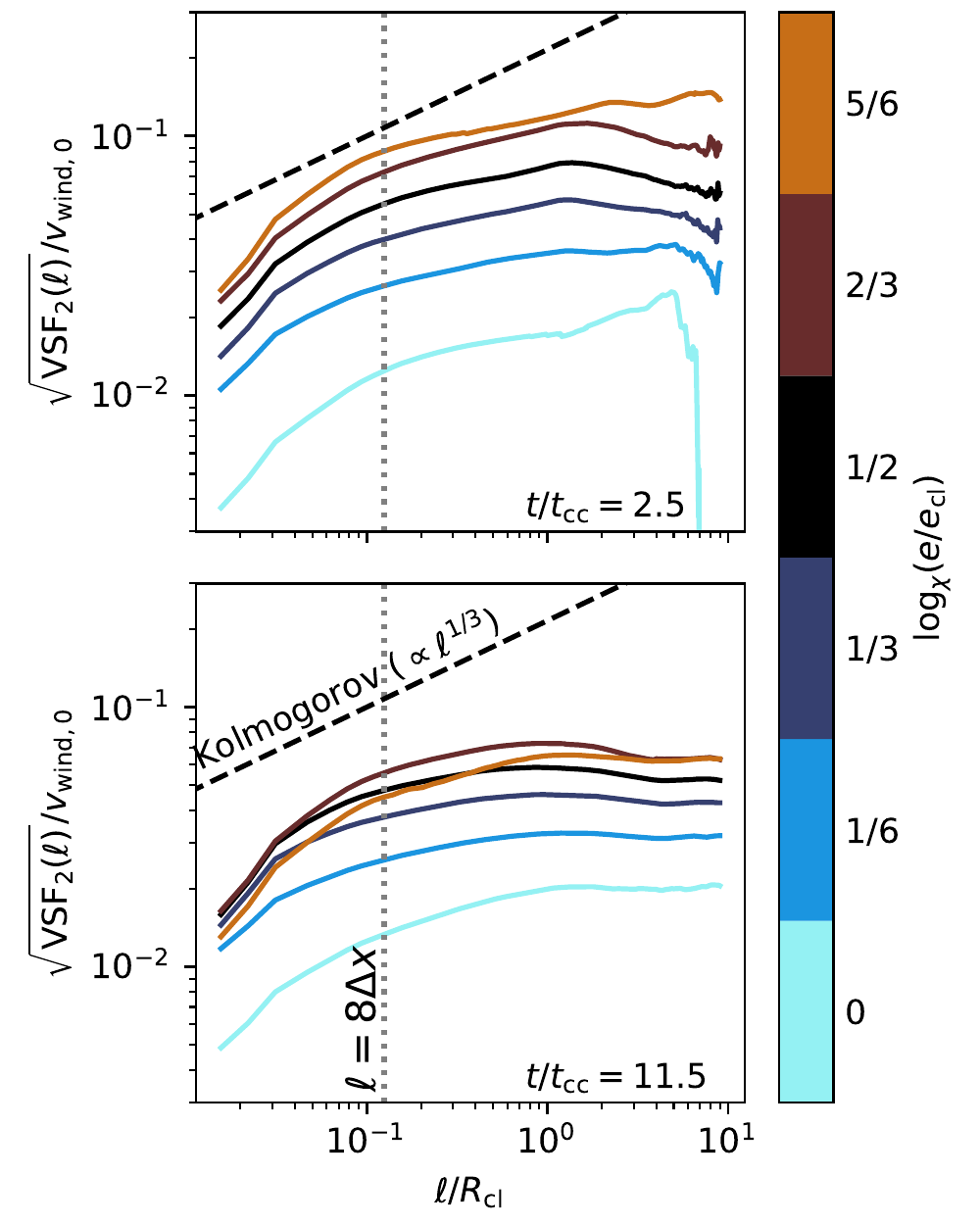}
  \caption{\label{fig:sf_example} Shows $\sqrt{\vsfsecond}$ measured for each phase bin of the $\rcl/\cellwidth=64$ run of our $\chi = 1000$, $\xish = 27.8$, $\Mw=1.5$ simulation at \fidtime\ (top) and $11.5{\tcc}$ (bottom).
  The black dashed line indicates the expected slope for idealized Kolmogorov turbulence.
  Data to the left of the gray vertical dotted line ($\ell<8\cellwidth$), lies outside of the inertial range.
  } 
\end{figure}

Our final turbulence measure is the velocity structure function, which has the advantage of explicitly exploring the dependence on length scale $\ell$, but comes with some uncertainty due to the potential influence of gradients in the large-scale bulk flows. 

To compute this measure, we consider a velocity vector field that is sampled at a collection of points.
Let $|\delta v|$ denote the magnitude of the velocity difference between a pair of points.
We define the first and second order velocity structure functions, \vsffirst\ and \vsfsecond, as the average values of $|\delta v|$ and $(\delta v)^2$ for all pairs of points separated by a distance $\ell$.
Except where otherwise noted, the velocity differences only include components orthogonal to the wind direction (see ~\autoref{appendix:early-time-vturb} for further explanation).

Given the obvious phase dependence in our other turbulence metrics, we compute the structure function for individual phases of the gas in our simulations, using the same bins defined in \autoref{sec:hipass}.
All structure function calculations in this work are computed using all pairs of points from individual phase bins. 
We note that both points in each pair always comes from the same phase bin, and we leave consideration of cross-phase terms to future work.
We omit the hottest phase-bin from our analysis because a large fraction is laminar (contaminating the signal) and it is computationally expensive to compute.

We also use discrete bins of $\ell$, which depend on the cell width, $\cellwidth$, in our simulations.
The $i$th $\ell$ bin is centered on $\ell = i \cellwidth$ and has a width of $\cellwidth$. 
However, for $i=0$ and $i=1$, we have adjusted the bins such that they only contain values for pairs of cells that share a face and an edge, respectively.
In other words, the $i=0$ bin ($i=1$ bin) only contains values for cells exactly separated by $\ell=\cellwidth$ ($\ell=\sqrt{2}\cellwidth$).

Throughout this work, we largely focus on $\sqrt{\vsfsecond}$ because
it has a similar magnitude to our other $\vturb$ metrics.
The top panel of \autoref{fig:sf_example} shows $\sqrt{\vsfsecond}$ measured for each phase bin of our $\chi = 1000$, $\xish = 27.8$ simulation at \fidtime.
The peak in \vsfsecond\ at $\ell \sim \rcl$, present in all phases (in some cases it manifests as a change in slope), is expected since the outer scale should be of order the cloud size \rcl; although the complicated cloud structure at late times is unlikely to correspond to a narrow range for the injection of turbulence.
We leave investigation of the behavior above \rcl\ to future work.

We generally observe a weaker dependence on $\ell$ than the $\propto \ell^{1/3}$ scaling expected for idealized, Kolmogorov turbulence (for $\sqrt{\vsfsecond}$), although this depends somewhat on phase.
We caution that the precise $\ell$ scaling at intermediate (inertial) scales may not be a robust measurement due to the bottleneck effect, which arises for under-resolved turbulent cascade \citep[e.g.][]{Rennehan:2021,mohapatra22_filaments}.

In agreement with the other measures, we also see a general decrease in the turbulent velocity with temperature.

\subsection{Comparison}\label{sec:turb_comp}

\begin{figure*}
  \center
\includegraphics[width = 7in]{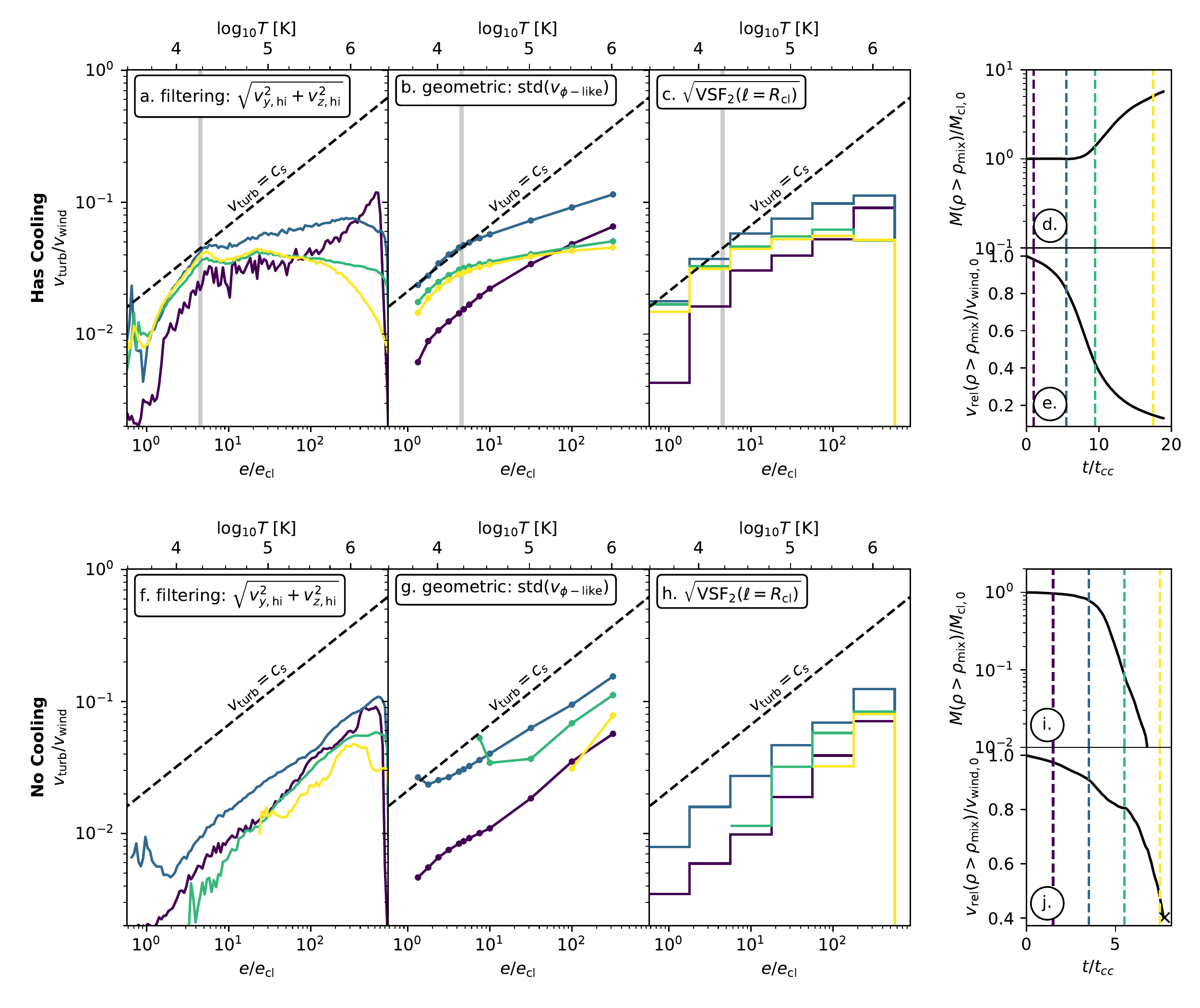}
\caption{\label{fig:metric-comp-turb-phase}
  Each row shows phase-dependence of \vturb\ measured with different metrics (large panels), and bulk property evolution (smaller panels) for a $\chi=1000$, $\Mw=1.5$, $\cellwidth = \rcl/16$ simulation.
  The top row shows our $\xish = 27.8$ simulation, (the cloud is entrained) and the bottom row shows the same simulation without cooling (the cloud is destroyed).
  The \vturb\ panels respectively show data measured with high-pass filtering, from isosurfaces, and using \vsfsecond.
  The data is colored by the time at which they are measured (the vertical lines in the small panels denote the times) and the dashed black line indicates values of \cs\ as a function of phase (assuming constant pressure).
  Note that the lowest temperature point on the blue and green curves of the isosurface panel in the lower row, are likely outliers: the relevant isosurfaces probably bound a small amount of mass.
  The bulk evolution panels respectively show the mass in the cold phase, (i.e. gas denser than $\sqrt{\rhocl\rhow}$) and the average velocity difference between the cold phase and \vw\ as functions of time.
}
\end{figure*}

We have shown that much more information about the phase, scale, and spatial dependence of turbulence can be gleaned from these simulations when using metrics beyond the standard root-mean-square approach. We now compare these more refined turbulent metrics to each other.

The top row of \autoref{fig:metric-comp-turb-phase} shows the \vturb\ phase dependence, measured with each approach, in a $\rcl/\cellwidth = 16$ of the run of the previously mentioned simulation at $t = 1.0, 5.5, 9.5, 17.5 \tcc$ (see \autoref{sec:convergence} for a discussion of how resolution affects our measurements).

The differing approaches achieve remarkable qualitative agreement about the magnitude, phase dependence, and temporal dependence of \vturb. In particular, all measures show that \vturb\ increases rapidly with temperature at early times, before transitioning to a flatter slope at later times.
In addition, all approaches show very similar amplitudes.
However, the approaches are clearly not interchangeable.
Indeed, this plot demonstrates the utility of computing all three turbulence measures, allowing us to ascertain the robust results without over-interpreting features that are not seen in at least two of the techniques.

When considering the volume-averaged properties of the entire system, our geometric approach offers the most robust measurements because it is most resilient to biases that may arise from the gradients in the laminar component of the flow at early times (see \autoref{appendix:early-time-vturb} for more details).

In the context of phase-dependence, the filtering approach clearly is the most convenient metric because it naturally provides turbulence as a function of phase.
However, unlike the other approaches, the filtering approach does not examine the turbulence of one phase in isolation to the others, which may introduce ``artifacts'' in this type of comparison.
We will show in \autoref{sec:convergence}, that the negative slope at large $T$, at late times may be a resolution effect.

The $\sqrt{\vsfsecond}$ approach captures much of the same phase dependence while also opening a window into the scale dependence of the turbulence.
With that said, it is the most computationally expensive.

\section{Results} \label{sec:results}

\begin{figure*}[htb!] %
  \center
\includegraphics[width = \textwidth]{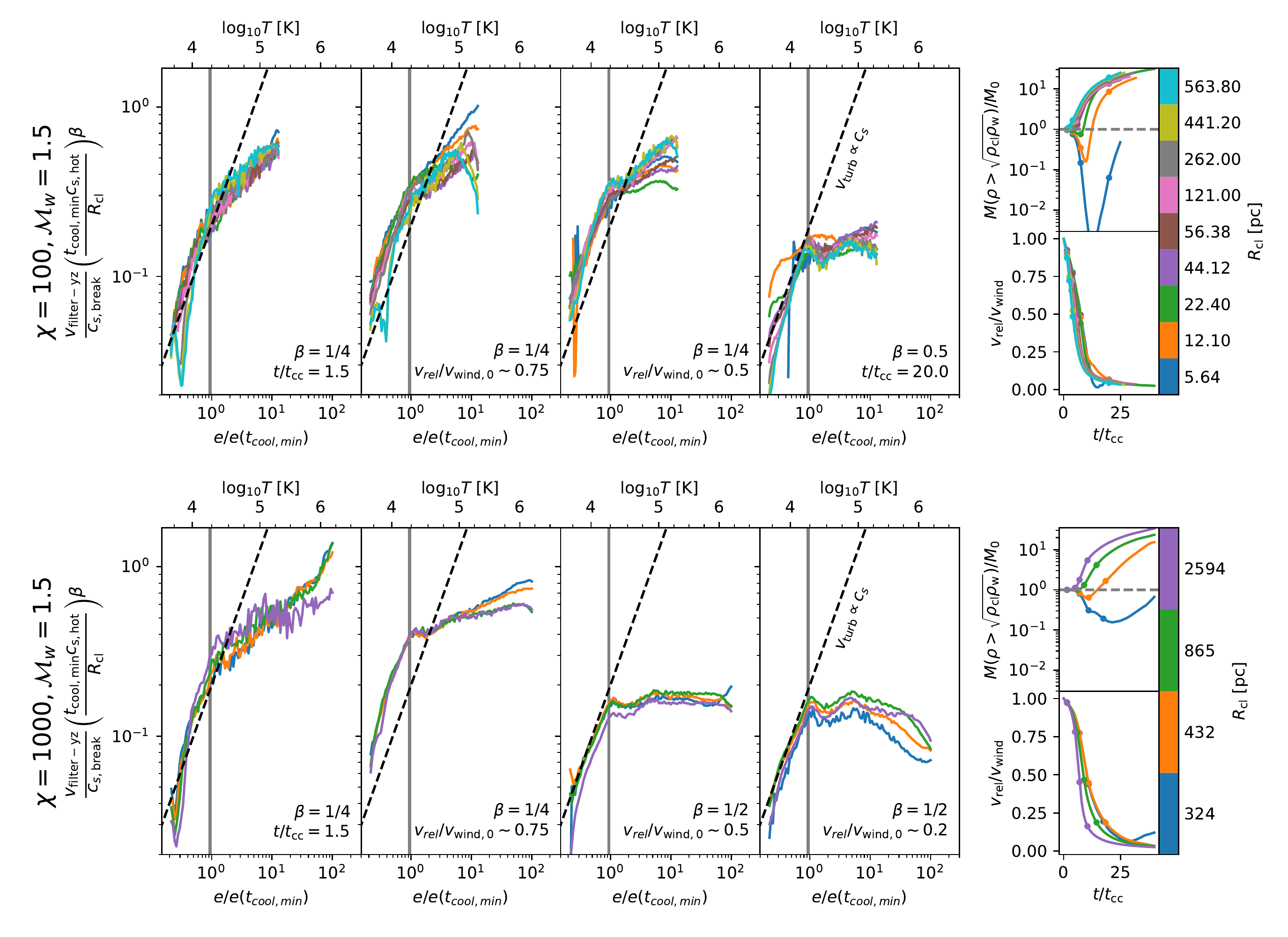}
\caption{\label{fig:rcl-dependence}
    The top row compares the dependence of \vturb\ with gas phase at four representative times in the clouds' evolutions (full-size panels on the left) and bulk property evolution (small panels on the right) for a separate collection of simulations.
    The top row compares 9 simulations with $\chi=100$, $\Mw=1.5$, and the same initial cloud temperature, but varying \rcl\ (and so varying \xish).
    The \vturb\ panels shows filtering measurements after $1.5{\tcc}$ (leftmost panel), when the relative velocity between the cold and hot phase are various fractions of its initial value (middle 2 panels), and after $20{\tcc}$ (right panel).
    Data is only shown for a given simulation for $0<\logXe<0.9$.
    The black dashed line shows $\vturb=\cs\propto\sqrt{e}$ and the vertical grey line extends between the temperatures where the cooling length is minimized and $\tcool$ is minimized.
    The bulk property panels show the evolution of (upper) the cold phase's mass and (lower) the relative velocity in each simulation.
    The curves in these panels are annotated with dots to specify the snapshots displayed in the \vturb\ panels.
    It's a little ambiguous whether the cloud survives in the $\chi=100$, $\tshear\sim0.57$ run, or if its destruction seeds the prompt precipitation of cold phase material (see \autoref{tab:sim_table} for more details).
    The bottom row shows the same information, but for a set of 4 simulations that all have $\chi=1000$.
    The other difference is that the rightmost \vturb\ panel in the bottom row compares \vturb\ measurements at $\vrel/\vw\sim0.2$. 
    We note that $\cshot\tcoolmin$ is 6.56 pc (20.7 pc) for simulations in the top (bottom) row.
}
\end{figure*}

\begin{figure*}
  \center
\includegraphics[width = \textwidth]{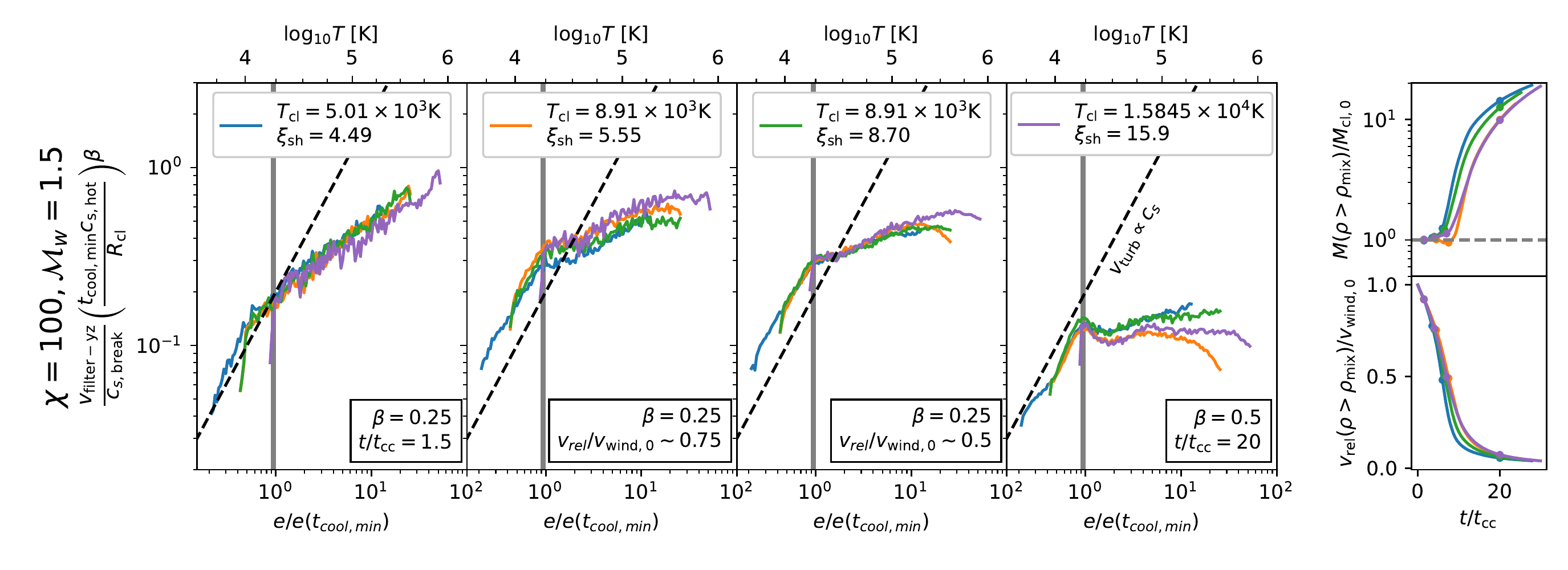}
\caption{\label{fig:vturb-Tcl-dependence}
    Like the top row of \autoref{fig:rcl-dependence} except that the pictured simulations primarily vary the cloud temperature.
    Each simulation has $\chi=100$ and $\Mw=1.5$.
    We expect at higher resolution that the power-law slope below \ebreak\ in the purple curve will be closer to 0.5 (i.e. the slope of the dashed black line).
  }
\end{figure*}

\begin{figure*}
  \center
\includegraphics[width = \textwidth]{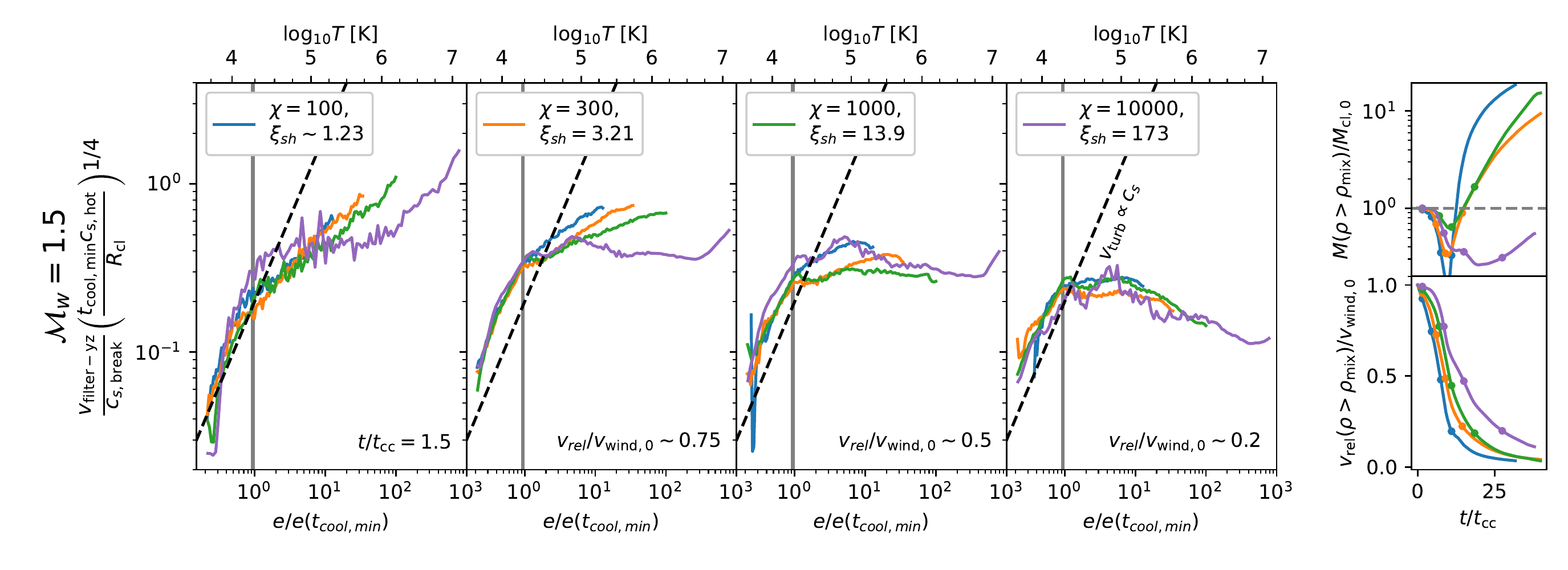}
\caption{\label{fig:chi-vturb-dependence}
    Like the top row of \autoref{fig:rcl-dependence} except that the pictured simulations primarily vary in $\chi$.
    We have made two compromises in presenting this data.
    First, we fix $\beta$ to $0.25$ for all panels.
    This is done as a simplification because $\beta$ changes on a timescale related to $\chi$.
    Second, the rightmost \vturb\ panel compares simulations at a fixed value of $\vrel/\vw$ rather than at a fixed time.
    The last panel typically compares the simulations at a point in evolution when \vturb\ stabilizes (see \autoref{sec:time_evo}).
    However, that time seems to come much later in our $\chi=10^4$ simulation, after the simulation terminates.
    While we include the $\chi=10^4$ run for completeness, strong resolution dependence (see~\autoref{tab:sim_table}) and the atypical shape of the cool-phase mass evolution may indicate that it is not well-converged.
    As noted in \autoref{sec:simulations}, some material that started in the cloud leaks out of the domain at $6.5{\tcc}$, which coincides with the large drop-off in cool-phase mass.
}
\end{figure*}

\begin{figure*}
  \center
\includegraphics[width = \textwidth]{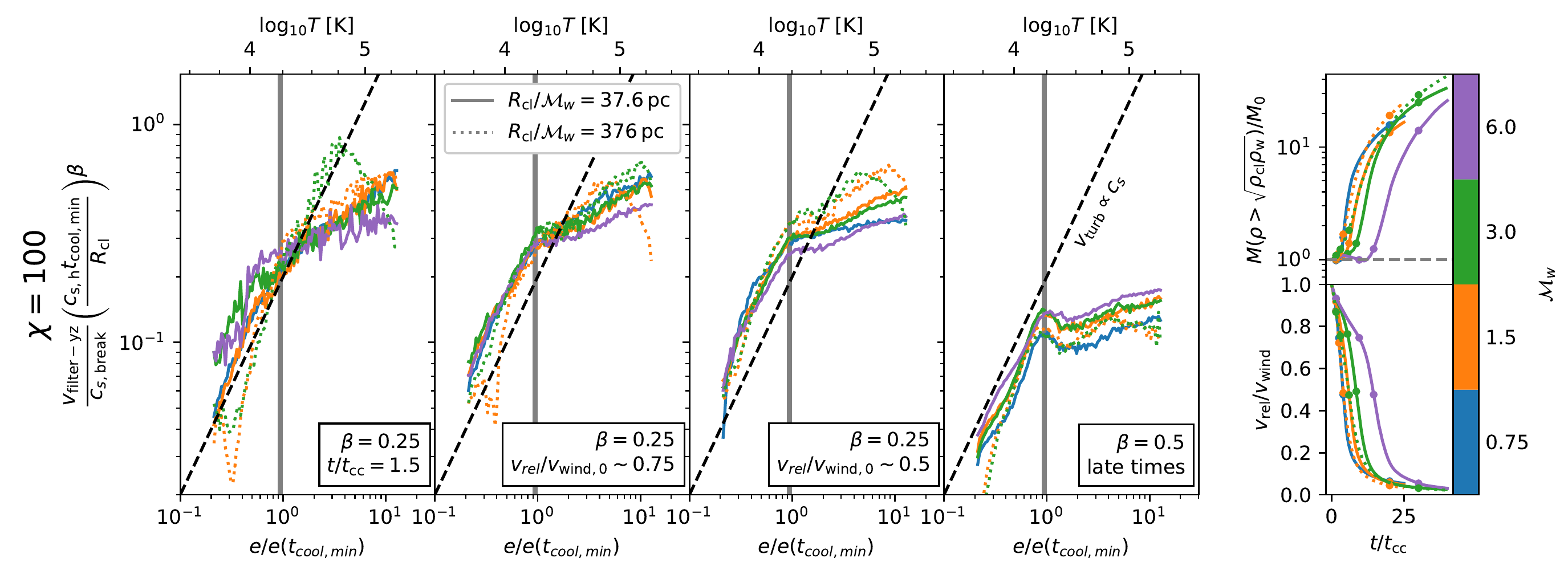}
\caption{\label{fig:vturb-mach-dependence}
    Like the top row of \autoref{fig:rcl-dependence} except that the pictured simulations primarily vary \Mw.
    The solid (dotted) lines show data from simulations with $\rcl/\Mw = 37.6\, {\rm pc}$ ($376\, {\rm pc}$) and $\xish = 5.73$ (57.3).
    We note that the $\cshot\tcoolmin$ is 6.56 pc for all simulations in this plot.
    As we will show in panels e-h of \autoref{fig:hysteresis_general}, \vturb\ evolves more slowly in higher \Mw\ runs.
    Consequently, the ``late times'' panel shows data from $\Mw=0.75,1.5$ runs at $20{\tcc}$, and data from $\Mw=3,6$ runs at $30{\tcc}$ (we did not run the $\Mw=6$ simulation to late enough times or with a long enough domain for an optimal late-time comparison).
  }
\end{figure*}

Having established the robustness and relative merits of our turbulence metrics, we now examine what they tell us about the cloud-wind interaction.
We start (in \autoref{sec:turbulent_properties}) by describing the phase dependence of \vturb, its scaling with dimensionless parameters, and time-dependence.
Then, in \autoref{sec:driving_scale} we briefly discuss the driving scale before turning to an evaluation of numerical convergence in \autoref{sec:convergence}.

For the purpose of this discussion and subsequent sections, we define the cold phase as all gas with densities of at least $\sqrt{\rhocl\rhow}$ (i.e. the density of the mixing layer).
We also define the relative velocity, \vrel, as the difference between \vw\ (at the inflow boundary) and the mass-weighted velocity of the cold phase (initially \vrel\ is \vw\ but declines as the gas is entrained).

\subsection{Turbulent Properties}\label{sec:turbulent_properties}

Throughout this group of subsections, we will compare simulations with different parameters and $\cellwidth=\rcl/16$.
We first consider the phase dependence of \vturb, then show how \vturb\ scales between simulations, and finally describe the time-evolution of \vturb.

\subsubsection{Phase dependence}\label{sec:phase_dependence}
We begin by presenting the phase dependence in two limiting cases of the $\chi=1000$ and $\Mw=1.5$ cloud-wind interaction.
These two cases are: (i) a run without cooling ($\xish=0$) and (ii) a run where cooling is sufficient for entrainment ($\xish=27.8$).

The bottom row of \autoref{fig:metric-comp-turb-phase} shows the non-radiative run.
In this case, the scaling of \vturb\ with $e$ (or $T$) is consistent with a power-law where $\Mturb$ is constant (i.e. $\vturb \propto \cs \propto \sqrt{e}$) throughout the cold phase's lifetime. 
The amplitude of the turbulence decreases as \vrel\ drops but there is no indication that the scaling with phase changes (note that at late times, the cold gas is entirely absent, due to mixing with the hot phase and so we cannot measure its turbulent properties).
As expected, these trends are unaffected by our choice of turbulent metric.

The top row of \autoref{fig:metric-comp-turb-phase} shows the run with cooling.
Compared to the constant-slope power-law phase dependence of turbulence in our non-radiative run, the case with cooling clearly has more complex behavior.
We parameterize the phase dependence of systems with sufficient cooling for entrainment, at a given time, using a broken power-law,
\begin{eqnarray}
\label{eqn:phase-dependence}
   \frac{\vturb(e)}{v_{\rm turb,break}} \sim  
\begin{cases}
   \sqrt{e/\ebreak} & \text{if } e \leq \ebreak\\
   \left(e/\ebreak \right)^\alpha & \text{if } \ebreak \leq e 
\end{cases},
\end{eqnarray}
with a break at $\ebreak=\emincool$, which coincides with the minimum of \tcool.\footnote{\emincool\ also roughly coincides with the location where \lcool\ is minimized (the thickness of the vertical gray lines in the top row of \autoref{fig:metric-comp-turb-phase} denotes the difference in locations), but this may not be the case for different physical conditions.
}
For now, we're just interested in $\alpha$; the following subsections will discuss $v_{\rm turb,break}$.

Below \ebreak, the scaling of \vturb\ on $e$  is constant in time.
Above \ebreak, the slope of the power-law dependence, $\alpha$, has clear time-dependence.
At very early times (${\la}\tcc/8$), geometric measurements provide some evidence (not shown) that $\alpha=1/2$; in this case \autoref{eqn:phase-dependence} is equivalent to the scaling of our non-radiative run.
As the system evolves, $\alpha$ decreases (i.e. the slope flattens above \ebreak).
When the cloud is mostly entrained, $\alpha$ stabilizes at $\sim0$.\footnote{
We show that negative values of $\alpha$ at late times are likely a resolution effect in \autoref{sec:convergence}.}
While we don't show it, we note that similar behavior occurs in our $\chi=1000,\xish=2.78$ run, but the cloud is destroyed long before $\alpha$ drops to 0.

This demonstrates an essential feature of the turbulence in systems with sufficient cooling for cloud survival and entrainment: \emph{the cold phase has a larger turbulent Mach number and turbulent kinetic energy than the hot phase}.

\subsubsection{Scaling with Cloud Parameters ($\chi$, \Mw, and \xish)}
Now that we've established the behavior in these limiting cases, we discuss how the \pdns\ ($\chi$, \Mw, and \xish) affect the magnitude of \vturb\ in simulations with rapid enough cooling to ensure cloud survival.
At a given stage of a cloud's evolution (i.e. for a given value of $\vrel/\vw$ or fixed time), we find that \vturb\ satisfies the scaling,
\begin{equation}
\label{eqn:vturb-mag-scaling}
    \frac{\vturb(\ebreak)}{\csbreak} \propto
    \left(\xish \Mw\right)^{\beta} \propto 
    \left(\frac{\rcl}{\cshot\tcoolmin}\right)^{\beta}, 
\end{equation}
where \xish\ and \Mw\ both refer to values used to initialize the problem.
This is equivalent to saying that \vturb\ scales with the ratio between the hot-phase sound-crossing time (\rcl/\cshot) and \tcoolmin.
The best fit values for $\beta$ are $0.25$ and ${\sim0.5}$ at early and late times, respectively.
This change in $\beta$ seems to coincide with a transition between temporal evolutionary stages, which we will discuss further in the next subsection and link to a change in the primary source of turbulent kinetic energy.

Figures~\ref{fig:rcl-dependence}-\ref{fig:vturb-mach-dependence} compare \vturb\  measurements, adjusted to remove differences captured by this scaling, for different sets of simulations.
In other words, the agreement between the curves in a given panel in these figures indicates the accuracy of the adopted scaling.
Because the \pdns\ clearly affect the slope of \vturb\ above \ebreak, the reader should primarily consider agreement at \ebreak\ (denoted by a vertical line) and in colder gas.
Note that unlike previous plots, the black dashed line shows $\vturb\propto\cs(e)$ rather than $\vturb=\cs(e)$.

First, we consider the scaling for runs with $\Mw=1.5$.
\autoref{fig:rcl-dependence} shows the scaling on \rcl; the top (bottom) row shows runs with $\chi=100$ ($\chi=1000$). 
The impressive overlap of the curves in each panel demonstrates that the adopted scaling works remarkably well -- there are occasional differences at high $e$, but the turbulence in the gas closest to the wind phase is the most challenging to accurately measure.
The figure also clearly shows that the shape of the turbulence dependence with $e$ changes over time, a point we will return to later.

Figures~\ref{fig:vturb-Tcl-dependence} and \ref{fig:chi-vturb-dependence} provides evidence that \vturb\ depends not just on \rcl, but on the ratio $\rcl/\cshot$ by comparing runs with different \rcl\ \emph{and} \cshot\ values.
The variation in \cshot\ come from adopting different values for \Tcl\ and $\chi$.
\autoref{fig:chi-vturb-dependence} provides further confirmation that \cshot\ is the correct sound-speed to include in this scaling because \cshot\ has different $\chi$-dependence from the sound speed in the (cold) cloud phase.

Finally, \autoref{fig:vturb-mach-dependence} demonstrates the \vturb\ scaling for runs that vary in \Mw. 
It largely confirms the lack of \Mw\ dependence.

We provide a rough normalization for \autoref{eqn:vturb-mag-scaling} when $\vrel/\vw\sim0.75$.
In this case, we find that $\vturb(\ebreak)\sim 0.4 \csbreak (\rcl/(\tcoolmin \cshot))^{1/4}$.
The precise normalization will change when using other techniques to measure \vturb.

All of these results are computed with the filtering metric for turbulence. We note that these scalings are somewhat less clear for geometric and \vsfsecond\ measurements of the $\chi=100$ simulations (the scaling between $\chi=1000$ runs is clear for all metrics).
For example, the geometric measurements show slightly different trends among the runs that initially lose mass, and suggest that $\beta$ never changes from 0.25 in the $\chi=100$ runs.
Although the latter quirk is difficult to explain, we are encouraged by the fact that the geometric approach does show the change in $\beta$ for the $\chi=10^3$ simulations, and the fact that the \vsfsecond\ relation definitely supports $\beta=0.5$ at late-times in our $\chi=100$ simulations.
With that said, \vsfsecond[\ell=\rcl] measurements do show more scatter than is present in the top row of \autoref{fig:rcl-dependence}.
While this could be physical, the coarser phase bins may also contribute to this scatter.
We defer further investigation to future work.

\subsubsection{Temporal evolution} \label{sec:time_evo}

\begin{figure*}
  \center

\includegraphics[width = \textwidth]{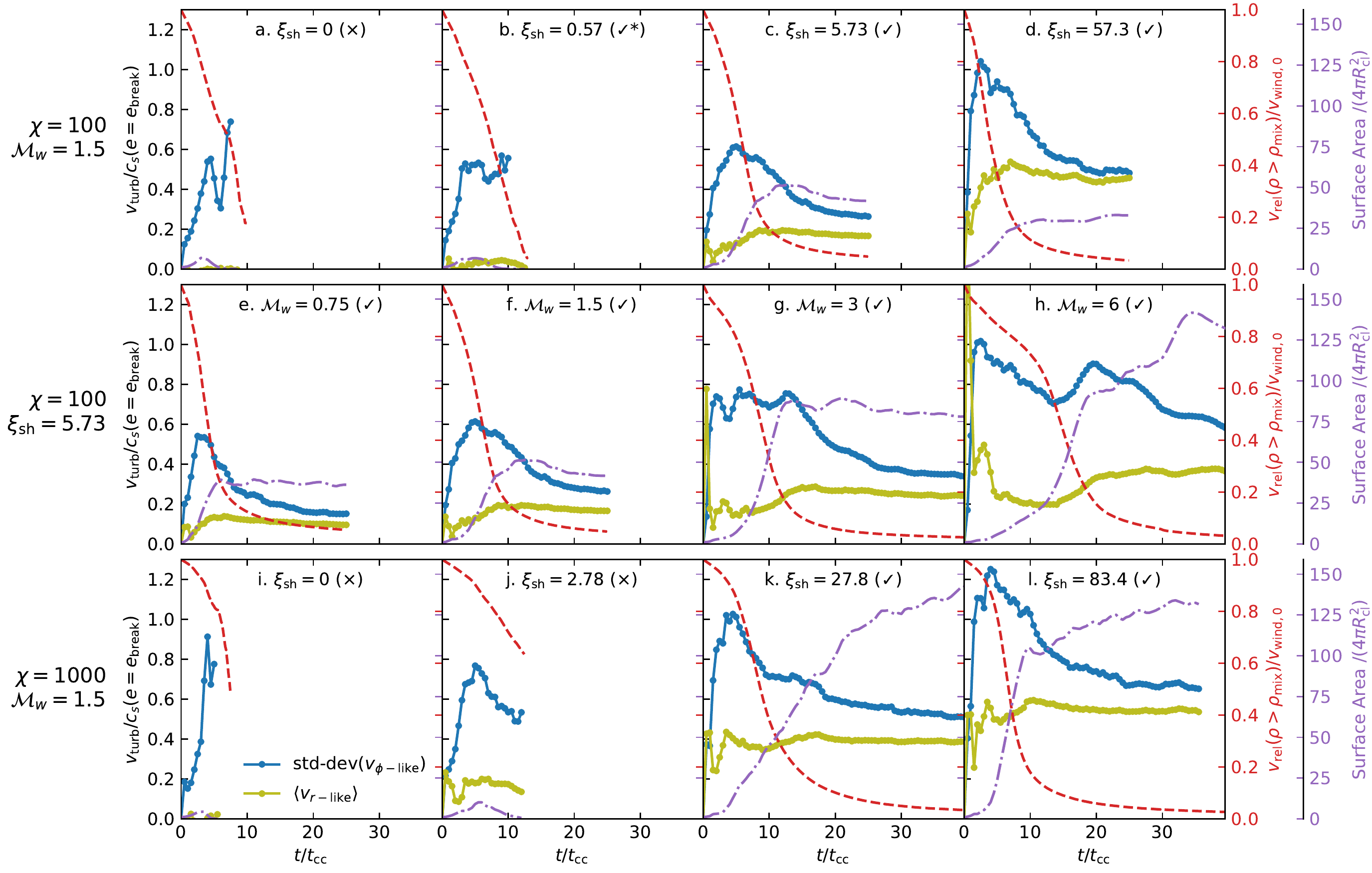}
\caption{\label{fig:hysteresis_general}
  Each panel shows the temporal evolution of \vturb\ (blue solid curve), the average radial inflow (olive curve), the surface area (violet dashed-dotted curve), and \vrel\ (dashed red curve).
  The \vturb, average inflow, and surfaces areas were all computed from an isosurface constructed at the temperature at which $\tcool$ is minimized (this is $e/e_{\rm cl} = 4.8$).
  More specifically, the average inflow is computed from the area-weighted average of the normal velocity component on each facet on the isosurface.
  In contrast, $\vrel$ measures the bulk relative velocity of all gas with $\rho > \sqrt{\rhocl\rhow}$ and is normalized such that it starts at unity and approaches zero as a cloud is entrained.
  To denote that a cloud becomes entrained (is destroyed) we include a ``$\checkmark$'' (``$\times$'') in the panel label.
  The top (bottom) row show runs that have $\chi = 100$ ($\chi = 1000$) and $\Mw=1.5$ while varying \xish.
  The middle row shows runs that have $\chi = 100$ and $\xish=5.73$ while varying \Mw.
  Panels c and f show the same run.
  As noted in the \autoref{tab:sim_table}, the cloud's fate in panel b is somewhat debatable, given that the cold-phase mass drops to 0.07\% of it's initial value before growing.
  For this case, we have elected not to show data after the cloud starts growing.
}
\end{figure*}

So far, we have focused on how the phase-dependence of the turbulence changes (at a set of different times) with cloud properties. 
We turn our attention to how \vturb\ changes with time in a single simulation.
Given how the slope of the \vturb\ broken power-law phase dependence is largely time independent for the cold phase up to the break, we focus on \vturb\ at $e=\emincool\sim\ebreak$.
\autoref{fig:hysteresis_general} shows, for a broad range of simulations, the time evolution of \vturb, measured geometrically (we use this metric to isolate a narrow phase bin), the average inflow velocity, and surface area on the same isosurface.
The figure also shows the time evolution of \vrel. %
We do not show other types of \vturb\ measures because they are less accurate at early times (see \autoref{appendix:early-time-vturb}), and do not distinguish between turbulence and gradients in inflowing gas as well as the geometric measurements.

We start by considering the turbulent evolution in a characteristic case with cloud entrainment: \autoref{fig:hysteresis_general}c shows a $\chi=100$ run with $\Mw=1.5$ and $\xish = 5.73$.
\vturb\ has two primary evolutionary stages.
Initially, in the `pre-entrained stage', \vturb\ rapidly grows until it reaches a peak value;
\vturb\ is sustained near this peak for a short time, and then it starts to drop off, as the cloud becomes partially entrained.
During the subsequent `partially entrained' phase, \vturb\ stabilizes at a smaller value (within a factor of ${\sim}2$ of the peak) that is sustained for the remainder of the run.

The primary source of turbulent energy during the pre-entrained stage appears to be the relative velocity.
This would explain why \vturb\ peaks within a few \tcc: we expect large \vrel\ to drive the Kelvin-Helmholtz and Rayleigh-Taylor instabilities, which have growth rates proportional to \tcc\ \citep{klein94a}.
This also explains similar rapid turbulent growth during the  initial stage of the non-radiative and the slow cooling simulations in panels a and b of \autoref{fig:hysteresis_general}.
Furthermore, it explains why the drop in \vturb, which indicates the transition between stages (and is most prominent in the radiative runs), follows the drop in \vrel\ -- this is presumably because \vrel\ no longer provides enough turbulent energy to sustain the peak \vturb.

The two stages of \vturb\ evolution roughly coincide with the stages of areal growth identified in \citet{gronke20a}.
The `pre-entrained' stage coincides with the rapid surface area growth dominated by the formation of the cloud's tail.
Likewise, the `partially entrained' stage roughly corresponds to the slower isotropic areal growth that occurs once the cloud is entrained.
It's also noteworthy that the average inflow velocity plateaus before the slower isotropic areal growth, which is consistent with findings from \citet{gronke20a}.

At face value it might seem surprising that there is net inflow in \autoref{fig:hysteresis_general}b even though we know that this run is losing mass during the first ${\sim}10{\tcc}$ (see the mass evolution of the $\chi=100$, $\rcl=5.64\, {\rm pc}$ run in \autoref{fig:rcl-dependence}).
However, this just illustrates that net inflow doesn't necessarily equate with mass growth; the inflowing gas will raise the temperature of the gas enclosed by an isosurface in the absence of sufficient cooling.

We now consider how the \pdns\ (\xish, $\chi$, \Mw), affect the \vturb\ evolution with time.
In general, we find that these parameters only minimally affect the overall trend, so we focus on the relatively small differences that do emerge.

First, we examine variation in \xish.
Compared to panel~c of \autoref{fig:hysteresis_general}, panel d illustrates that more efficient cooling can increase the maximum \vturb\ as well as the value of \vturb\ at late times.
This is consistent with the scalings from the last subsection.
In this panel, \vturb\ approaches \vinflow's magnitude at late times.
It's plausible that all entrained runs in the figure would show the same behavior if we had measurements for late enough times; it may just be most prominent in panel d because \vinflow\ is elevated and the cloud is accelerated more quickly.
This feature may suggest that \vturb\ is dominated by the radial flow at late times.
We also find that higher \xish\ simulations have a somewhat smaller surface area.

The bottom row of \autoref{fig:hysteresis_general} shows data for a set of runs with $\chi=10^3$, and varying entries of \xish.
In simulations in which the cloud survives, the transition between evolutionary stages of $\vturb$ happens at larger $\vrel/\vw$ when $\chi$ is larger. 
This transition appears to roughly coincide with the time at which the value of $\beta$, from \autoref{eqn:vturb-mag-scaling}, increases from $0.25$.
Differences in \vturb's magnitude are qualitatively consistent with the scaling given in that equation.

Finally, the middle row of \autoref{fig:hysteresis_general} compares runs with varying \Mw.
Increasing \Mw\ appears to increase \vturb's initial growth rate, \vturb's magnitude, and the duration over which \vturb's maximum magnitude is sustained.
There is also some indication that higher \Mw\ simulations may also have larger inflow rates and larger surface areas, even at late times.

\begin{figure}
  \center %
  \includegraphics[width = 3.35in]{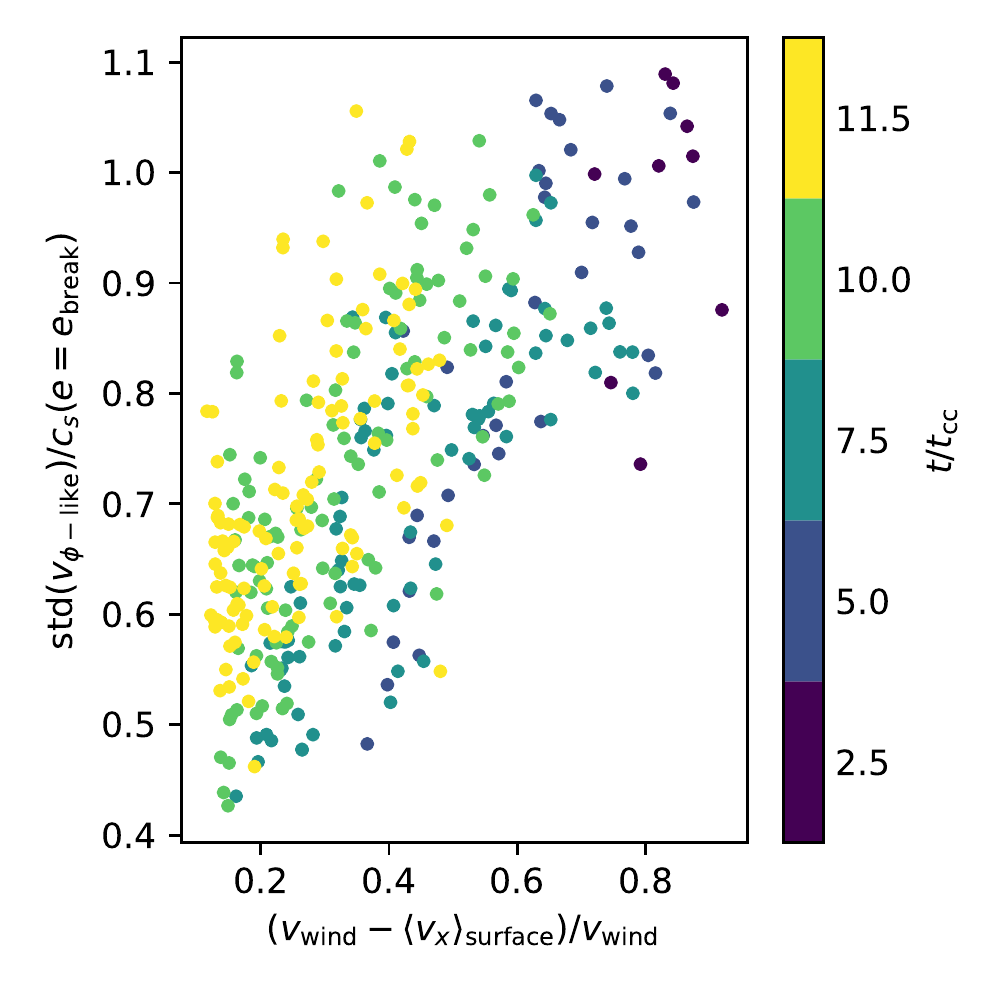}
  \caption{\label{fig:diff_flow} Points of a given color show \vturb\ and $(\vw - v_{\rm isosurface})/\vw$ measurements for different sections of the \ebreak\ isosurface from $\rcl/\cellwidth = 64$ run of our $\chi=10^3,\xish=27.8$ simulation.
  The points' colors indicate the simulation time that the measurement is associated with.
  The isosurfaces are split into bins based on each facet's position along the \vwindhat.
  Each bin has a width of \rcl; there are more bins when the cloud is more elongated.
  The averages and standard deviations are all weighted by the area of each facet.
  While it is not shown, we have evidence indicating the data's slope may change when the \pdns\ are varied.
  }
\end{figure}

Independent of \xish, $\chi$, and \Mw, \autoref{fig:hysteresis_general} illustrates that the acceleration timescale is tightly correlated with the stages of areal growth (the surface area and \vrel\ curves feature abrupt slope changes at similar times).
In contrast, the transition between \vturb\ stages appears less tightly coupled with the acceleration timescale as the \pdns\ are changed.
We attribute this mostly to the fact that the cold phase is not a rigid body with a single bulk velocity, but instead has different velocities at different spatial locations.

This differential acceleration is responsible for the cloud's head-tail morphology: downstream material moves faster than upstream material.
Regions with larger \vrel\ (compared to \vw) should generally have larger \vturb, albeit with some scatter related to the local history of turbulent driving.
This is illustrated for a high resolution version (to improve sampling) of our $\chi=1000,\,\xish=27.8$ run in \autoref{fig:diff_flow}. Here, we explore the relation between our measured turbulence metric (at the cooling peak) and the relative velocity of the gas as a function of both time (colors) and location along the length of the cloud (different points with the same color). This demonstrates that there is a correlation between these quantities not just at different times for the whole cloud (as shown in \autoref{fig:hysteresis_general}), but also along a cloud at a given time, strengthening the case for a causative relation.

How does this relate back to the loose coupling seen between the \vturb\ evolutionary stages and the acceleration timescale, when we vary the \pdns?
Because entrained clouds in our various simulations have different wind-aligned lengths, we know that changes in these numbers alter the cloud's differential acceleration.
Consider the temporal evolution of the volume-averaged \vturb\ measurements for a narrow phase bin of a very coherently accelerated cloud and a less coherently accelerated cloud.
One would naturally expect that that \vturb\ measurements might spend more time near its maximum value in one of these cases.
It's not much of a stretch to assume that \vrel\ might be fairly different when \vturb\ starts to decrease (i.e. begins transitioning between stages).
Thus, we would find different coupling between \vrel's evolution and \vturb's evolution in these cases.

\subsection{Evolution of the driving scale} \label{sec:driving_scale}

We now briefly revisit the velocity structure function in order to investigate how the turbulent driving scale varies with time.
The bottom panel of \autoref{fig:sf_example} shows $\sqrt{\vsfsecond}$ for the $\rcl/\cellwidth=64$ run of our $\chi=1000,\ \Mw=1.5,\, \xish=27.8$ simulation when the cloud is mostly entrained in the wind ($\vrel/\vw\sim0.27$).
Comparisons with the top panel ($\vrel/\vw\sim0.94$) reveal that the outer scale of turbulent driving, which coincides with the peak \vsfsecond, does not change substantially from early to late times.
Although we don't show it, we confirmed similar behavior in the $\rcl/\cellwidth=32$ run of our $\chi=100,\ \Mw=1.5,\ \xish=5.73$ simulations for similar values of \vrel\ and at times when the cloud is more entrained.

We note that it's unclear why the $9/12 \leq e/\ecl < 11/12$ phase bin's \vsfsecond[\ell\sim0.3\rcl] measurement, from the lower panel, is smaller than comparable measurements for other phase bins.
This feature also appears in the $\rcl/\cellwidth=32$ version of this simulation.
In contrast, this feature is absent from the aforementioned $\chi=100$ run; in that case \vsfsecond\ is always larger for a given $\ell$ in hotter gas.

\begin{figure*}
  \center
\includegraphics[width = \textwidth]{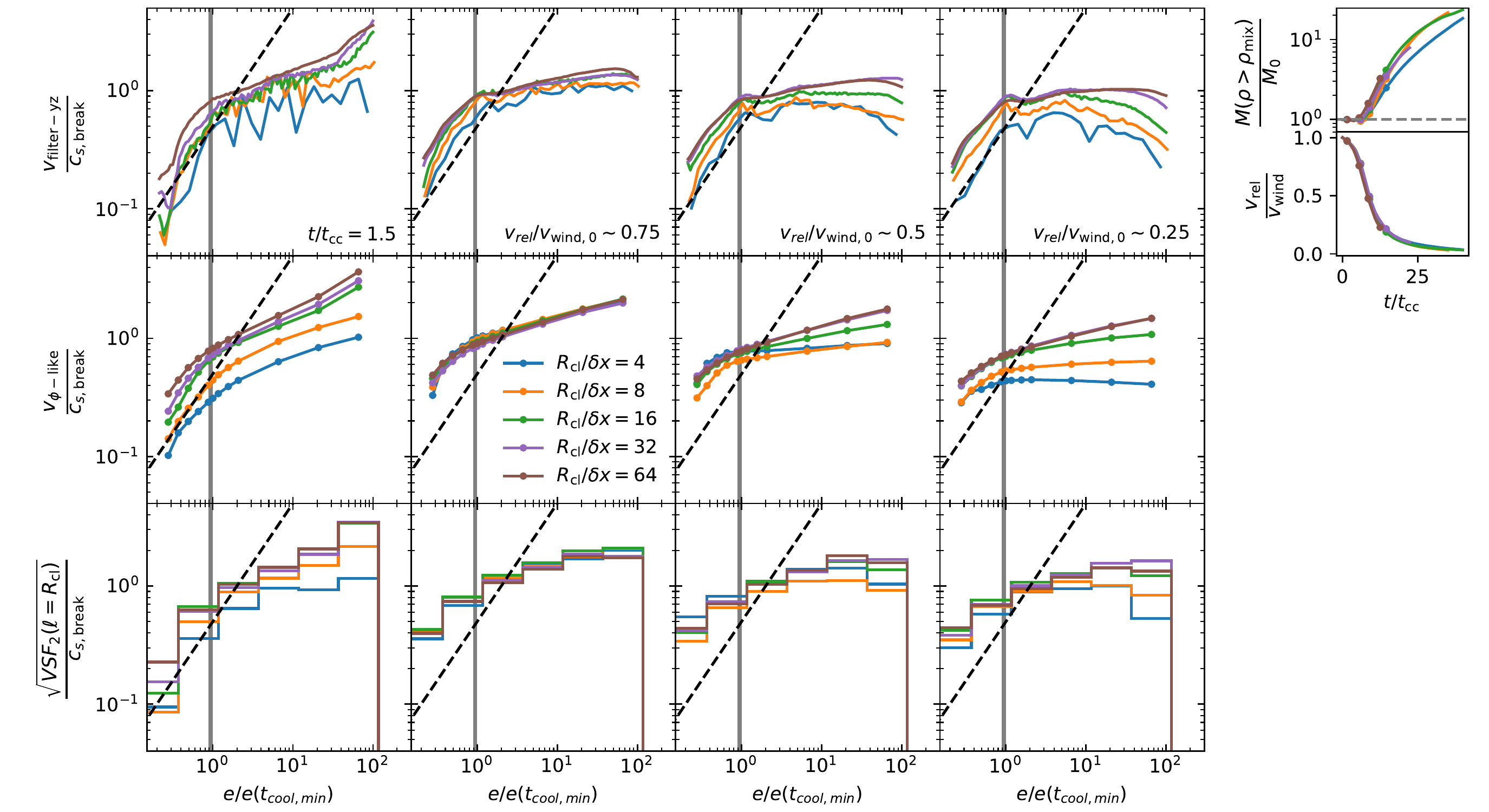}
\caption{\label{fig:vturb-res-dependence}
    The top row compares the \vturb\ phase dependence of different resolution runs of a simulation at 4 selected stages of evolution (full-size) and bulk property evolution (small panels) of each simulation.
    The illustrated simulations all have $\chi=10^3,\, \xish=27.8,\, \Mw=1.5$.
    The \vturb\ panels shows filtering measurements after $1.5{\tcc}$ (left panel) and when the relative velocity between the cold and hot phase are various fractions of its initial value (other panels).
    Data is only shown for a given simulation for $0<\logXe<0.9$.
    The black dashed line shows $\vturb=\cs\propto\sqrt{e}$ and the vertical grey line extends between the temperatures where the cooling length is minimized and $\tcool$ is minimized.
    The bulk property panels respectively show the evolution of (top) the cold phase's mass and (bottom) the relative velocity in each simulation.
    The curves in these panels are annotated with dots to specify the values during the snapshots displayed in the \vturb\ panels.
    The second and third rows of \vturb\ panels are the same as the top row, except that they respectively display geometric and \vsfsecond[\ell=\rcl]\  measurements.
    The $\rcl/\cellwidth=32,\, 64$ runs make use of a smaller simulation box than the other displayed runs.
    While visual inspection of our simulations, lead us to expect boundary effects to bias measurements in those cases, when $\vrel/\vw\sim0.25$, this doesn't seem to be an issue for this exercise.
}
\end{figure*}

\begin{figure}
  \center %
  \includegraphics[width = 3.35in]{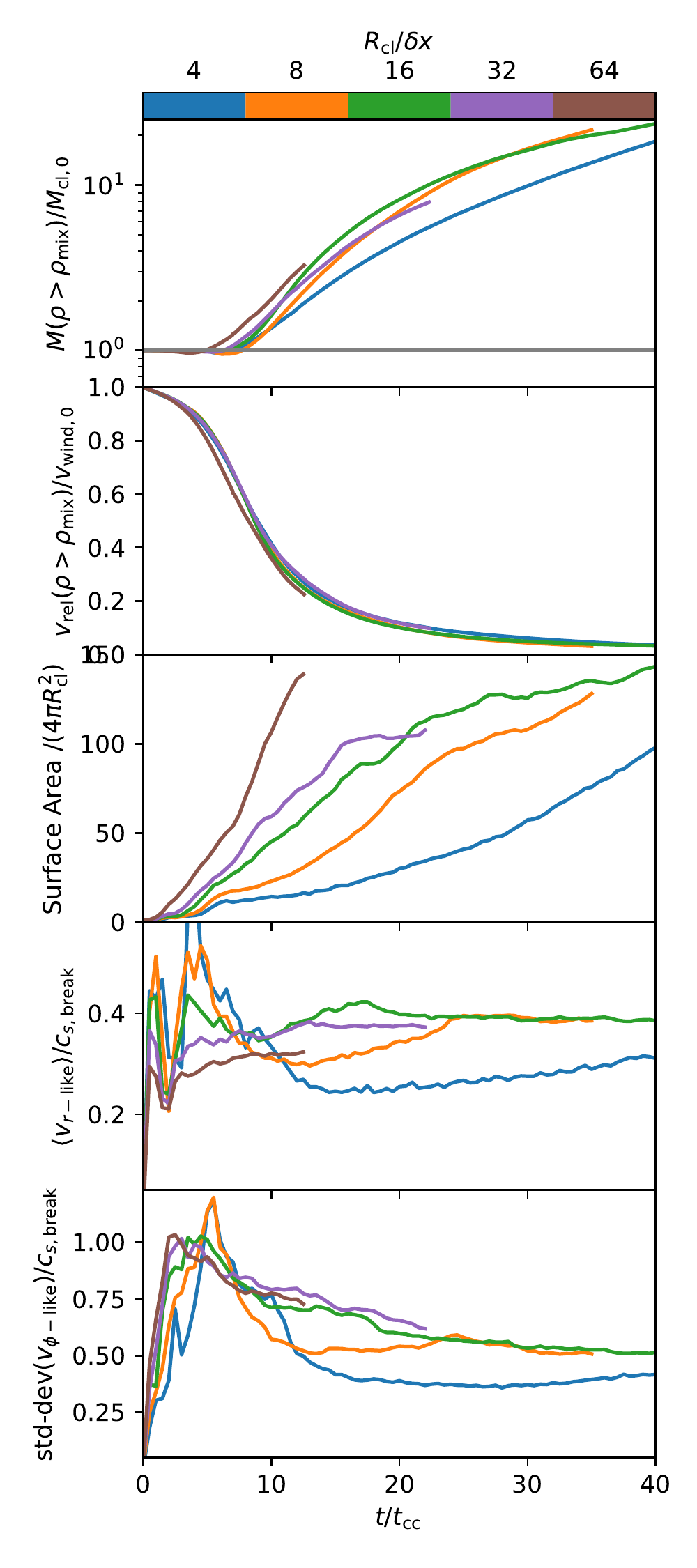}
  \caption{\label{fig:hysteresis_X1000_res}
    Illustrates how the evolution of various quantities in our $\chi=1000$, $\xish=27.8$ simulation are affected by resolution.
    The top two rows show evolution of the cold-phase mass and of the relative velocity between the cold and hot phases.
    Subsequent rows show evolution of quantities computed from the \ebreak\ isosurface including surface area, average inflow velocity, and the turbulent velocity.
  }
\end{figure}

\begin{figure}
  \center
\includegraphics[width = 3.35in]{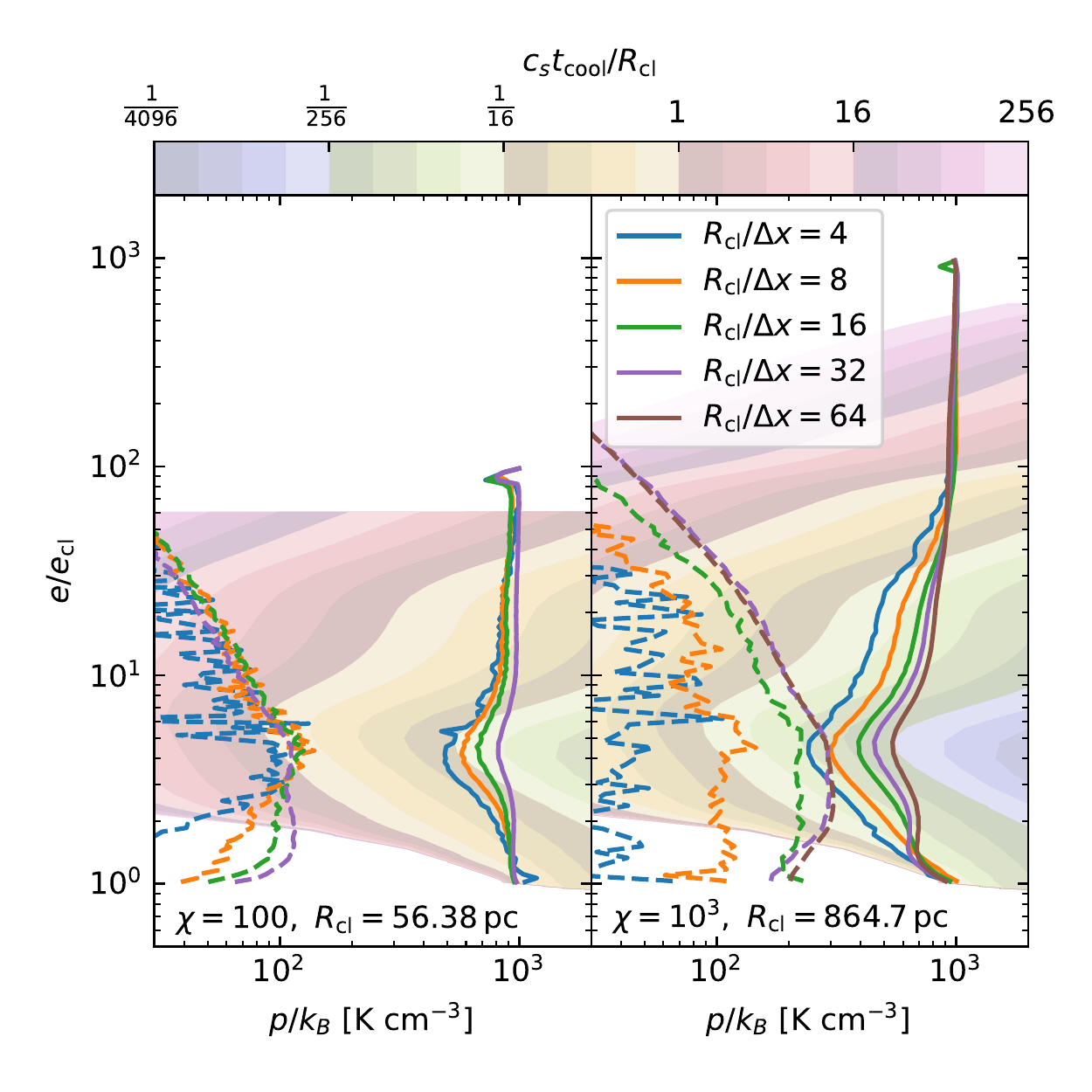}
\caption{\label{fig:phase-converge}
    Solid colored lines show the median thermal pressure as a function of temperature for multiple resolutions of our $\Mw=1.5$ simulations with $\chi=100,\xish=5.73$ (left) and $\chi = 10^3,\xish=27.3$ (right) at $11.5{\tcc}$.
    The background shaded region shows the size of the cooling length associated with a point in phase-space measured relative to the simulation's $\rcl$.
    There is not an associated length scale below \ecl\ or above ${\sim}\ew$ because we have disabled cooling and heating at these temperatures.
    At low pressures, just above \ecl\ there isn't an associated length scale because heating dominates.
    For the sake of comparison, the dashed lines show the median turbulent pressure, $\rho \vturb^2$ (derived from the filtering approach).
}
\end{figure}

\begin{figure*}
  \center
  \includegraphics[width = 0.85\textwidth]{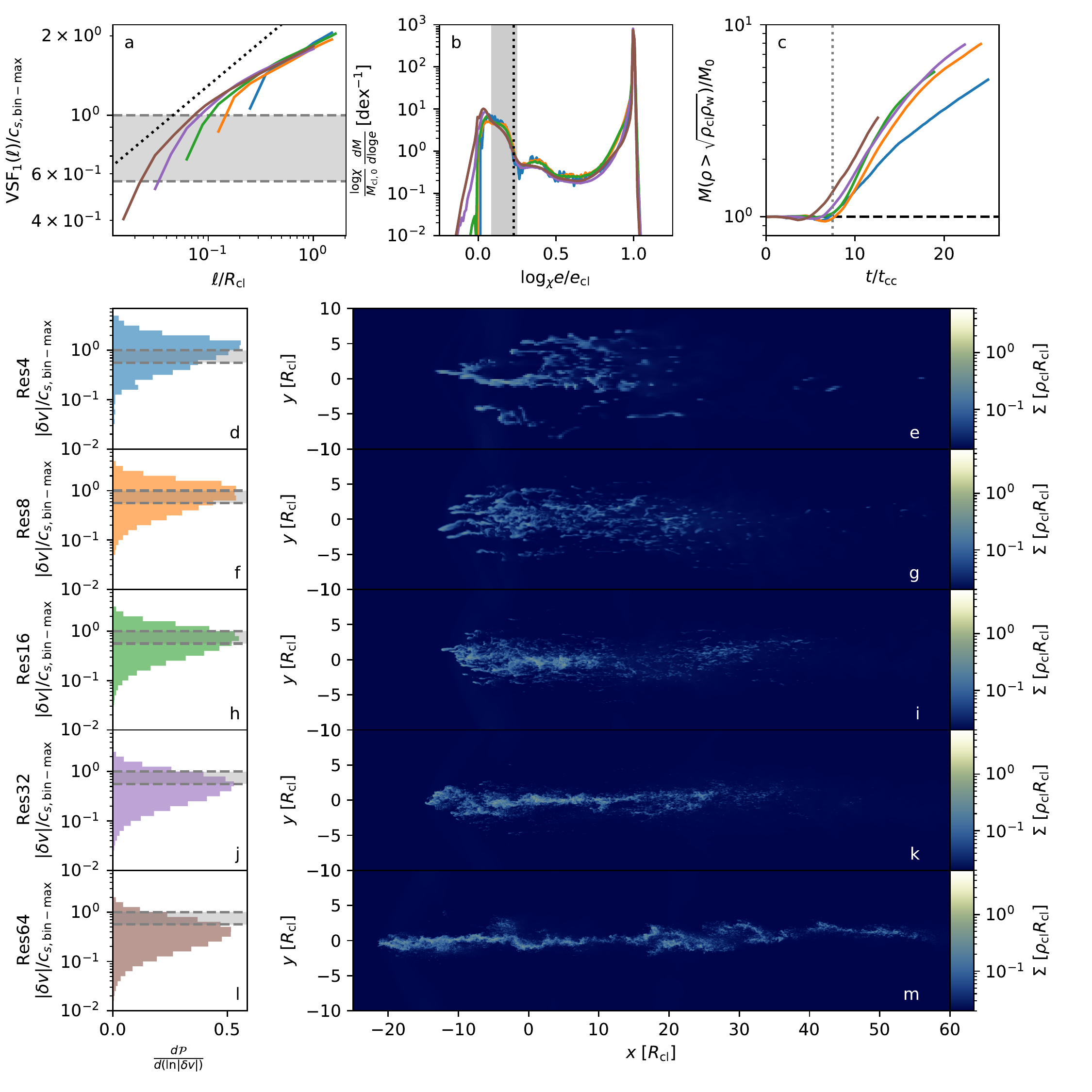}
  \caption{\label{fig:resolution_shattering}
    Illustrates resolution effects on the $\chi = 1000$, $\xish = 27.8$ simulation at $7.5{\tcc}$.
    The top row shows \vsffirst\ measurements for gas with $1/12 \leq \logXe < 3/12$ (panel a), the projected 1D phase distribution (panel b), and the bulk mass evolution for cold gas with $\rho>\rhomix$ (panel c).
    The dotted black line in panel a shows $\vsffirst\propto\ell^{1/3}$, the scaling expected for Kolmogorov turbulence.
    The brown shaded region in panel $b$ denotes the gas phases considered in \vsffirst\ while the vertical dotted line indicates the location of \tcoolmin.
    While the top row shows measurements from all resolutions, subsequent rows only show data for individual simulations.
    Panels d, f, h, j, and l shows the distribution of velocity difference magnitudes at the grid scale for gas with $1/12 \leq \logXe < 3/12$ (the average of this distribution is $\vsffirst[\ell=\cellwidth]$).
    The region enclosed by the grey dashed lines in these panels and panel a denote the range of \cs\ values for the selected phase bin.
    Panels e, g, i, k, and m shows the density projection for each run.
}
\end{figure*}

\subsection{Convergence} \label{sec:convergence}

In this section, we discuss how numerical resolution impacts our various measurements.
We primarily compare the measurements among different resolution runs of our $\chi=1000, \xish=27.8$ simulation, varying \rcl/\cellwidth\ from 4 to 64.

\subsubsection{Turbulence Metrics} \label{sec:convergence-metrics}

The large panels in the top row of \autoref{fig:vturb-res-dependence} compare the phase dependence of \vturb\ using filtering measurements of \vturb\ at various points in the cloud's lifetime.
The figure shows that resolution appears to slightly affect the magnitude and the slope of the \vturb\ phase dependence above \ebreak.
Importantly, the figure also suggests that the occurrence of a negative slope of the phase dependence is likely a resolution effect. 
The full phase dependence of $\vturb$ is well converged for $\rcl/\cellwidth \geq 32$.
These same conclusions apply to our other turbulence metrics (shown in the other rows).

\autoref{fig:hysteresis_X1000_res} shows how resolution affects the temporal evolution of various quantities.
The top panels show convergence in the total cold phase mass\footnote{As an aside, we do see some indications that resolution may strongly affect a cloud's fate in other simulations close to the survival threshold. However we defer further investigation to future work.}
and \vrel; the only noteworthy feature is that rapid growth begins slightly sooner at higher resolutions.
However, the surface area measurements are not converged at all; it increases more rapidly for higher resolution runs.
These results are consistent with the findings of \citet{gronke20a} for a $\chi=10$ simulation.

There are some differences in the \vturb\ evolution.
While low resolution runs have a strong, sharp peak followed by a flat region, higher resolution runs have a moderate peak with a gradual descent.
With that said, there seems to be convergence for $\rcl/\cellwidth\ga16$, and all of the runs qualitatively agree with our picture that there are two stages of \vturb\ evolution.
The average inflow velocity measurements are similar overall but do show some significant differences -- its somewhat unclear what the relevant trends are.
We defer further investigation of inflow velocity convergence to future work.

\subsubsection{Phase structure} 

Resolution strongly affects the 2D thermodynamic $e-p$ phase-space distribution.
Previous work \citep[e.g.][]{fielding20a,abruzzo22a} established that gas in $\chi\leq100$ simulations is roughly distributed along the isobar that is bounded by the properties of the cloud and wind.
However, a pressure decrement emerges in the phase distribution at points along this isobar where cooling is not resolved.

Each point in the internal energy-pressure ($e$-$p$) phase space has an associated cooling length-scale $\cs \tcool$.
\autoref{fig:phase-converge} shows that the size of the pressure decrement scales inversely with how well $\cs \tcool$ is resolved.
The figure also suggests that resolving the minimum cooling length scale (i.e., $\cellwidth\lesssim \lcool\sim \min(\cs\tcool)$), which is equivalent to the ``shattering'' length scale  \citep{mccourt18a}, is adequate to largely remove the pressure decrement for $\chi\sim100$, which is consistent with results from prior works \citep[e.g.][]{abruzzo22a}.

Under-resolved cooling is not the sole reason for the gas distribution's deviations from the pressure isobar.
\citet{ji19a} previously argued that it is actually the sum of the turbulent pressure, $\rho\vturb^2$, and thermal pressure that should match the external pressure.
\autoref{fig:phase-converge} illustrates the median turbulent pressure as a function of $e$ with dashed lines.
In both $\chi$ cases, the turbulent pressure shows clear convergence in our higher resolution runs.
The turbulent pressure's lack of dependence on $e$ below \ebreak\ and inverse correlation with $e$ above \ebreak\ (at the pictured time) are consistent with the scaling described in \autoref{eqn:phase-dependence}.
The factor of ${\sim}3$ difference in the maximum values (i.e. at $e\sim\ebreak$) of the turbulent pressures between the two $\chi$ cases helps explain why the $\chi=1000$ case has larger deviations in the thermal pressure from the external pressure. 
This difference is consistent with the scaling from \autoref{eqn:vturb-mag-scaling}.
For context, we expect $\vturb^2$ at \ebreak\ to be a factor of ${\sim}4.85^{2\beta}$ larger in this $\chi=1000$ case, although the value of $\beta$ is ambiguous; \autoref{fig:rcl-dependence} suggests that these particular $\chi=1000$ and $\chi=100$ runs should have $\beta=0.5$ and $\beta=0.25$ at $11.5{\tcc}$.

At the finite resolutions of our simulations, there is a decrement in the total pressure in our simulations.
However, at infinite resolution it is plausible that the total pressure of the gas is constant.
In short, the minimum $\cs\tcool$ along the segment of the pressure isobar, connecting the cloud and the wind phase properties, specifies the grid-scale requirement for fully resolving phase-structure.
Remarkably, the degree to which we resolve \lcool\ appears to have minimal impact on the 1D $e$ phase distribution.
This is shown in \autoref{fig:resolution_shattering}b (we will discuss the rest of the figure in the next section).

\subsubsection{Turbulent structure and Cloud Morphology} \label{sec:convergence-lturb}

Our results hint that under-resolving turbulence might influence various properties of these interactions.
To illustrate this, we turn to \autoref{fig:resolution_shattering}, which shows measurements taken from different resolution runs of our $\chi=10^3,\xish=27.8$ simulation at $7.5{\tcc}$.
Each panel in the top rows displays lines of data that comes from each resolution.
Subsequent rows just show measurements taken from a single resolution.

\autoref{fig:resolution_shattering}a shows the first-order velocity structure function, \vsffirst, measured for gas in the $1/12 \leq \logXe < 3/12$ ($8.1\times 10^3\, {\rm K} \leq T \leq 2.0\times 10^4\, {\rm K}$) phase bin at various resolutions.
Values are divided by the bin's maximum sound speed and the gray region denotes the width of the bin.
\vsffirst\ specifies the average magnitude of the velocity differences\footnote{Unlike for our \vsfsecond\ measurements, these calculations use the three-dimensional velocity vectors} for pairs of points separated by a length-scale $\ell$.

As the separation $\ell$ decreases so does the velocity difference.
On scales comparable to the cloud radius, $\ell \sim \rcl$, the slope and normalization of \vsffirst\ are remarkably well converged.\footnote{
The slope and normalization are less-well converged at earlier times ($t\la 4.5{\tcc}$).
} 
As the separation approaches the grid scale the velocity differences are damped by numerical dissipation. Where this numerical dissipation kicks in relative to the sound speed appears to have a major impact on the morphology of the system.
In reality the true physical viscosity of these systems is uncertain, but is likely to be much less than the effective numerical viscosity even in our highest resolution simulation.

On large scales the velocity differences are greater than the sound speed, but at small enough separations the velocity differences become subsonic. 
We define the turbulent sonic length, \lturbsonic, as the scale at which \vsffirst\ passes through the point $\vsffirst[\lturbsonic] = \cs$.
By extrapolating the slope from large separations we can estimate \lturbsonic\ in the limit of infinite resolution (and very small viscosity), which, in this case, falls around $0.07 \rcl$.
This \lturbsonic\ is not resolved by the simulations with $\rcl/ \cellwidth = 4$ or 8, is marginally resolved by the $\rcl/ \cellwidth = 16$ simulation, and is fairly well resolved by the $\rcl/ \cellwidth = 32$ and 64 simulations.
When $\cellwidth > \lturbsonic$, the average velocity difference, in a given phase bin, can be supersonic at the viscous scale (i.e. between adjacent cells).

Panels d, f, h, j, and l show the distribution of velocity difference magnitudes, in the previously mentioned phase-bin, measured at $\ell=\cellwidth$; the average values of these distributions give the leftmost points of the curves in panel a.
These panels illustrate that as $\lturbsonic/\cellwidth$ decreases, fewer pairs of cells have supersonic velocity differences.
They also show that some grid-scale supersonic velocity differences persist when \lturbsonic\ is barely resolved.

We now investigate the question: \emph{What are the consequences of under-resolving \lturbsonic?}
Panels e, g, i, k, and m of \autoref{fig:resolution_shattering} show the projected density of these simulations. The dramatic differences in these maps suggest that the degree to which \lturbsonic\ is resolved may be linked to morphological differences between simulations.
We find that the cold phase in higher resolution simulations is composed of more large-scale structures and has a narrower transverse extent, whereas in lower resolution simulations the cold phase is clumpier and more dispersed. The cold phase in the low resolution simulations has effectively \emph{shattered} while in the higher resolution simulations that have $\lturbsonic/\cellwidth > 1$ the cold phase remains more intact \citep{mccourt18a}.
These effects support a picture in which under-resolving turbulence intensifies shattering by enabling the presence of supersonic velocity differences on the grid scale.
This also naturally explain the wider dispersal of cold gas in low resolution simulations since the most intense shattering cause explosive breakup of clouds \citep{gronke20b}. Physically, supersonic grid-scale velocity differences will lead to large pressure imbalances that will in turn promote the dispersal as opposed to coagulation of cold cloudlets \citep{gronke22b}.

Using \autoref{eqn:vturb-mag-scaling}, which captures how \vturb\ scales with $\vrel/\vw$ and $\tcoolmin$, we can write a rough scaling relation for \lturbsonic.
Assuming that $\vsffirst\propto \ell^\zeta$, we find for cold gas with $\ecl\leq e \leq \ebreak$ that
\begin{equation}
    \label{eqn:lturbsonic}
    \frac{\lturbsonic}{\rcl} \propto \left(\frac{\tcoolmin}{\tshear}\right)^{\beta/\zeta}
    \Mw^{-1/(4\zeta)}.
\end{equation}
For sake of convenience we take $\zeta=1/3$ (the scaling for Kolmogorov turbulence), which is close to what is found in the simulations (see \autoref{fig:resolution_shattering}a). 
At early times in the `pre-entrained' stage (e.g., when $\vrel/\vw\sim0.75$) $\beta = 1/4$, which yields a precise prediction for the turbulent sonic length 
\begin{equation}
    \label{eqn:lturbsonic_pre_entrained}
    \left(\frac{\lturbsonic}{\rcl}\right)_{\rm pre-entrained} \approx 1.2 \left(\frac{\cshot \tcoolmin}{\rcl}\right)^{3/4} %
\end{equation}
The normalization is measured empirically in our $\chi = 1000$, $\xish = 27.8$ simulation. 
Note that we focus on the \vsffirst\ measurements from the same phase-bin that includes \ebreak\ since, as we have shown above, this is the region of phase space where these scalings are robust, but the general trends will be the same for other bins below $\ebreak$.
Due to the fact that we have only measured the length where \vsffirst\ is equal to the maximum sound-speed of the phase bin, this relation should be considered an upper-limit on $\lturbsonic$.

It's more intuitive to compare this relation against other known length-scales, like the minimum radius for cloud survival, $\rclcrit$ or the minimum cooling length. For fixed cloud properties, we find that $\lturbsonic/\rclcrit \propto \tcoolmin^{3/4} p^{3/4} \ecl^{-1.5} \chi^{-1.2}\Mw^{-1.7}$. \footnote{
This assumes that \rclcrit\ has the scaling from the \citet{li20a}/\citet{sparre20a} survival criterion, since this does an accurate job predicting cloud survival (see \autoref{sec:survival-criterion}).
Survival criteria have the generic form, $\tau_{\rm cool}/\tcc < q$ and thus $\rclcrit \propto \tau_{\rm cool} \Mw / q$.
In this case, $\tau_{\rm cool}\sim\tcoolw$ and $q\propto \rcl^{0.3} \Mw^{-2.5} n_{\rm w}^{0.3} \vw^{0.6}$, or equivalently $q\propto \Mw^{-1.9} \rcl^{0.3} p^{0.3} \mu_{\rm w}^{0.3}$.
For $T\ga 10^5\, {\rm K}$, $\tcool$ roughly scales as $e^{2.7}p^{-1}$ and the mean molecular weight, $\mu$, is constant.
Putting this together yields $\rclcrit^{1.3} \propto \chi^{2.7}\ecl^{2.7} \Mw^{2.9} p^{-1.3} \cscold$ when $\Tw\ga10^5\, {\rm K}$.
}
This demonstrates that the turbulent sonic length tends to be more difficult to resolve in runs with larger $\chi$ and higher $\Mw$.
If we assume that $\lcool=\min(\cs \tcool)\sim \csbreak\tcoolmin$ and $\csbreak\sim\cscold$, we find that $\lturbsonic/\lcool \sim \xish^{1/4}\Mw^{1/4}\sqrt{\chi}$.
Given our table of simulations, it should be clear that \lturbsonic\ exceeds \lcool\ in all of our entrained runs.

We find that this relation reproduces the value of \lturbsonic\ measured from the $\rcl/\cellwidth\sim32$ run of our $\chi\sim100, \Mw=1.5, \xish = 5.73$ simulation, to within ${\sim}50\%$.
The lower resolution runs of that simulation all resolve \lturbsonic, and we are encouraged that none of them shows signs of shattering (the transverse extent is fairly consistent among runs).
In the $\rcl/\cellwidth\sim16$ run of our, $\chi\sim 10^4, \xish= 172.8$ simulation, we find that \lturbsonic\ is smaller than the grid scale, when $\vrel/\vw\sim0.75$, which is consistent with the relation's prediction.
We note that both resolution runs of this simulation clearly shatter.
We performed a few spot-checks with a handful of our other runs and the relation seems accurate to within a factor of a few, but more careful modeling is required since $\lturbsonic$ is close to $\rcl/8$ and $\rcl/16$ in many of our runs.

While we primarily presented this analysis for the phase bin containing \ebreak, we also find evidence (not shown) suggesting that the general results can be extrapolated to lower temperature bins.
This is intuitive, given our earlier finding that $\vturb(e)/\cs(e)$ is roughly constant for $\ecl\leq e\leq\ebreak$.

Although our association of these large-scale morphological changes with \lturbsonic\ requires further investigation, it presents an attractive way to understand several outstanding related questions, namely, when do clouds shatter \citep{gronke20b}, and why do higher $\Mw$ simulations require so much higher resolution to achieve convergence \citep[][we elaborate further in section~\ref{sec:compare-prior-work}]{gronke20a,bustard22a}.
Although this need not be true in the general case (e.g. if there are external drivers of turbulence), $\lturbsonic > \lcool$ in all of our runs.
Thus, the resolution effects on large-scale morphology may be more closely related to under-resolved cooling.

We clarify that resolving small scale structure (e.g. surface area and number of clumps) has other conditions unrelated to resolving \lturbsonic.
\citet{sparre18a} and \citet{gronke20a} each show that convergence of such properties is very weak in high resolution simulations that resolve \lcool\ (in both studies, $\lturbsonic>\lcool$).

\section{Discussion} \label{sec:discussion}

\subsection{Phase Dependence of turbulence} \label{sec:discussion-phase}

We have demonstrated for the first time that the turbulent velocity, $\vturb$,  in a mixing layer follows a broken power law dependence on temperature or internal energy. 
A major implication of this finding is that the turbulent kinetic energy density is \textit{not} constant across gas phase.
Consider the ratio of the turbulent kinetic energy densities in the hot and cold phases, or 
$\epsilon = \rhow \vturb(\Tw)^2 / (\rhocl \vturb(\Tcl)^2)$.
Per \autoref{eqn:phase-dependence}, this evaluates to $\epsilon = (\ebreak/e_{\rm w})^{1-2\alpha}$.
We remind the reader that $\alpha$, the power-law slope above \ebreak, starts out near $1/2$ at the earliest times and decreases to ${\sim}0$ at a rate that depends on the \pdns.
Thus, during the bulk of the cloud-wind interaction, the cold phase has a larger turbulent kinetic energy density (i.e. $\epsilon<1$).
This contradicts (explicit and implicit) assumptions that $\epsilon=1$ in multiple works on TRMLs.

For example, we consider the arguments that lead to the expression for the temperature of the mixing layer, $\Tmix\sim\sqrt{\Tcl \Tw}$ \citep{begelman90a, gronke18a}.
This relation derives from the average of the cold and hot phase temperatures, weighted by the mass flux from each phase into the mixing layer.
The derivation assumes that each phase’s mass flux scales with the respective \vturb\ values.
Because the derivation involves arguments equivalent to assuming $\epsilon=1$, it overestimates the hot phase’s \vturb\ and consequently the mass flux when compared against the values for the cold phase.
Thus, $\sqrt{\Tcl\Tw}$ overestimates \Tmix\ and the size of the discrepancy is inversely correlated with $\epsilon$.
Because the value of $\tcool$ is commonly monotonic between \tcoolmin\ and $\tcool(\sqrt{\Tcl\Tw})$ \citep[e.g. see figure 14 of][]{abruzzo22a}, typical calculations overestimate \tcoolmix\ by an amount also negatively correlated with $\epsilon$.

In another case, \citet{fielding20a} explicitly assumes that $\epsilon=1$.
The only practical implication is that their quoted measurement of $f_{\rm turb} = \vturb/\vrel$ is too large by a factor of $\sqrt{\epsilon}$.
Thus, $f_{\rm turb}$ might have a weak dependence on the shape of the cooling curve.
In their analysis of clouds in a turbulent medium, \citet{gronke22a} also assumes $\epsilon=1$, but this may be valid since they consider externally driven turbulence.

\subsection{Observable Predictions}

It may be possible to observe \vturb's broken power-law phase dependence in real-world systems.
For example, previous studies have already placed constraints on temperature and nonthermal motion in the circumgalactic medium of other galaxies by measuring the widths of absorption lines for elements with different atomic masses \citep[e.g.][]{rudie19a,qu22a}.
Similar measurements may also be possible for high velocity clouds, for which there an abundance of absorption \citep[e.g.][]{fox04a} and emission line data \citep[e.g.][]{tufte98a, hill09a}.
One could also imagine using 21 cm emission or Mg\textsc{ii} absorption to extend such an analysis to probe the turbulent properties down to lower temperatures, where gas is atomic \citep[e.g.][]{marchal21a}.

It may also be possible to perform a similar exercise for %
gas in multiphase galactic outflows \citep[e.g.][]{Strickland:2009,ReichardtChu:2022}.

Additionally, one can perform more straight-forward comparisons against observational measurements of turbulent measurements in ${\sim}10^4\, {\rm K}$ gas.
However, given the simplifying assumptions in this work (described further in \autoref{sec:caveats}) and the fact the drivers of turbulence may vary between different systems, such comparisons must be interpreted with great caution.
Nevertheless, we find it encouraging that there is evidence that the Perseus molecular cloud has transonic turbulent Mach number \citep{burkhart15a}, just like we see in a fair number of our simulations.
We also find it encouraging that studies of CGM clouds \citep[e.g][]{rudie19a,qu22a} recover non-thermal broadening measurements within a factor of a few of $10\, {\rm km}\, {\rm s}^{-1}$, which nicely matches the turbulent velocities in our simulations.
We leave further comparisons to future work.

\subsection{What drives mixing?}

We now return to one of the motivating questions, the origin of turbulence in the flow.
From the results in this paper, the short answer appears to be that both shear and cooling drive the turbulence responsible for mixing.
As we conclude in \autoref{sec:phase_dependence}, shear is the primary driver of turbulence at early times.
After the cloud becomes partially entrained, \vturb\ falls off before stabilizing at a lower value.
The long-term support of a non-zero \vturb\ value, as \vrel\ goes to zero, suggests that some form of ``cooling-induced mixing'' mechanism takes over. 
To put this another way, the primary source of turbulent kinetic energy changes with time.
At early times, turbulent kinetic energy primarily comes from the large relative shear velocities between fluid elements.
At late times, it instead comes from the radial kinetic energy of inflowing material.

Possible origins for the late-time turbulence include rapid cooling driven pulsations in the cloud, \citep{gronke20a}\footnote{We did not save snapshots frequently enough to test our simulations for their presence.}, or simply the net radial inflow driven by the initial shear-driven turbulence. This later explanation is supported by the correlation of \vturb's late-time magnitude with \vinflow, which itself correlates with a run's cooling efficiency.
We plan to provide a detailed analysis of the temporal evolution of \vturb\ and its dependence on \vrel\ in a follow up work.

A few other features are consistent with this conclusion.
First, we find the rapid growth of surface area, when shear primarily drives mixing, and subsequent stabilization at a roughly constant value, when mixing is primarily driven by pulsations or radial inflow, to be consistent.
Second, the minimal variance in the driving scale, as the cloud is elongated, is also consistent.
At early times the driving scale is linked with the length of the wind-aligned axis of the cloud, of order \rcl.
Because the cloud's transverse extent doesn't change much with time, the typical radial separation between opposite inflow `fronts' of the clouds should still be of order $\rcl$ at late times.
Finally, the saturation of the inflow velocity after cooling-driven mixing has fully developed fits into this picture since the shear-driven contribution will have become subdominant.

\citet{gronke20a} noted that the anti-correlation between the cold cloud mass growth rate and \vrel\ might suggest that shear-driven turbulence from the \KH\ instability might not fuel mass growth, and instead might be a competing destructive process.
However, our most efficiently cooling $\Mw=1.5$ runs with $\chi=100,300,1000$ have significant \vrel\ when they start monotonically growing.
In other words, mass growth at early times in these runs should primarily arise from shear-driven turbulence.
With that said, mass growth is still negatively correlated with \vrel\ since the surface area is still increasing.

The evolution in the \vturb\ phase dependence is also consistent with this picture.
When shear primarily drives turbulence at early times, turbulent kinetic energy is roughly constant with phase (as in non-radiative simulations where shear is the only turbulent driver).
In contrast, when cooling drives turbulence, it does so primarily in regions with short cooling times, which explains why turbulence in the hot phase drops off.

\subsection{What is the mixing timescale?}

The canonical estimates for the characteristic mixing timescale are \tcc\ and \tshear.
We find that the turbulent velocity scales with $\rcl^\beta\cshot^{-\beta}\tcoolmin^{-\beta}$, where $\beta$ is 0.25 at early times and 0.5 at late times.
Notably, it has no dependence on \Mw\ for most of the cloud's evolution.
Therefore, the characteristic mixing time has no \vrel\ dependence.

With that said, the initial value of $\Mw$ \emph{does} affect the temporal evolution of \vturb.
\autoref{fig:hysteresis_general} also provides some indications that the magnitude of \vturb\ may have some dependence on \Mw\ at very early times.
Comparing panels g to h (as well as f to g) reveal that the peak values of \vturb, when $\vrel>0.8$, is larger in the higher \Mw\ run by more than the factor of $\sqrt{2}$ expected by \autoref{eqn:vturb-mag-scaling} from differences in $\rcl$.

\subsection{Survival Criterion} \label{sec:survival-criterion}

There has been great interest in the literature about the minimum radius for cloud survival \citep[e.g.][]{gronke18a, li20a, sparre20a, kanjilal20a, abruzzo22a, farber22a}.
We will provide more firm conclusions about this topic in an upcoming work \citepAbruzzoSurvival.
However, we do note that the our results are most consistent with the predictions of \citet{li20a} with the corrections described by \citet{sparre20a} for supersonic winds.

\subsection{Convergence}

What does it mean to resolve the cloud-wind interaction?
The obvious ideal is to achieve point-wise convergence, but this is generally prohibitively computationally expensive except in rare cases \citep[e.g.,][]{lecoanet16a}.
Short of this ultimate goal there are lesser gradations of convergence that depend on the question at hand. 
The easiest quantity to achieve convergence in is the net mass growth of the cold phase.
We show in \autoref{fig:resolution_shattering}c that the mass growth is fairly well converged for resolutions of $\rcl / \cellwidth \gtrsim 8$.
This likely corresponds to some minimum threshold to resolve any turbulent mixing, and is consistent with previous findings \citep[e.g.][]{gronke20a}. 
The hardest quantity to achieve converge in is the 2d $p-e$ phase distribution, which requires resolving the minimum cooling length ($\lcool$; also known as the shattering length).
Therefore, if one is interested in simply capturing the total amount of mass in the cold phase then the resolution requirements are much less onerous than if one is interested in capturing the detailed phase structure (or cloud morphology).
The details of the phase structure can be extremely important for comparisons to observations since the pressure decrement that develops in under-resolved simulations occurs in precisely the region traced by commonly observed ions, such as Mg\textsc{ii} \citep[e.g.,][]{Nelson:2021, Burchett:2021}.

Here we propose an intermediate convergence criterion for the large-scale morphology of cold structures which requires resolving the turbulent sonic length $\lturbsonic$ by several cells.
This is in general less stringent than the requirement to resolve the minimum cooling length.
At face value, the difficulty of resolving \lturbsonic\ in galaxy-scale simulations suggests that the detailed morphological properties of cool (${\sim}10^4\, {\rm K}$) gas, involved in TRML entrainment, within galactic outflows and the circumgalactic medium are unlikely to be correct.
However, the implications of accurately capturing the morphology may be more complex in more realistic systems because of the way cloud shape and size couples to other physical process absent in our simulations.
For example, in systems in which the hot phase is itself turbulent, such as in galactic wind simulations \citep[e.g.][]{schneider20a}, under-resolving $\lturbsonic$ may lead to artificially shattered clouds which will in turn be more likely to be destroyed than if they were able to remain coherent.
Therefore, having $\cellwidth < \lturbsonic$ may prove to be essential for determining the overall phase structure and evolution of turbulent multiphase flows that are ubiquitous in and around galaxies. 

This discussion about large-scale morphological convergence of cool gas in larger-scale models deserves elaboration on two finer points. 
First, it assumes applicability of our results about the emergent turbulent properties in the cloud-wind interactions; we discuss how the equilibrium \Tcl\ and shape of $\tcool(T)$ affect this in the next subsection (\autoref{sec:compare-prior-work}).
Second, we are extrapolating from simulations of isolated clouds, whereas larger-scale models often include multiple clouds in an outflow \citep[e.g.][]{cooper08a,kim18a,schneider20a}.
This is not an issue when the inter-cloud spacing is large enough for clouds to be treated individually, albeit with a hot phase that is already turbulent from upstream interactions.
However, more work is required to make predictions when the inter-cloud separation is small \citep[such work might use a multi-cloud setup akin to ][]{aluzas12a,banda-barragan20a}.

\subsection{Comparison to prior work}\label{sec:compare-prior-work}

At early times, when the \KH\ instability is the primary driver of mixing, one might expect similarities between our runs and the TRML simulations of \citet{fielding20a} and \citet{tan21a}.
Unfortunately, it's difficult to draw direct comparisons since those works highlight properties after reaching a quasi-steady state.
In contrast, our runs never reach such a state since \vrel\ evolves with time.
More meaningful comparisons could be made if the cloud was in a potential that was tuned to maintain \vrel\ at late times.
Additionally, \citet{tan21a} point out that we would likely expect different \vturb\ scaling to be dependent on geometry. 
Nevertheless, we find the presence of inflowing gas at early times to be encouraging (especially when juxtaposed with our adiabatic runs that don't have net inflow).
The fact that \vturb\ and the inflow velocity show signs of scaling with cooling efficiency is also encouraging.

Likewise, we expect similarities with \citet{gronke20a} at late times when turbulence is driven by “cooling-induced mixing” .
Although we broadly see similar qualitative evolution in the surface area, detailed comparisons of other properties are challenging.
While both works measured \vinflow, we expect differences in our methodologies will complicate comparisons of these quantities at late times.
\citet{gronke20a} used $\vinflow\sim\dot{m}_{\rm cold}/(A \rhow)$ while we directly measure the velocity component normal to the \emix\ isosurface (the scaling doesn't change much if we use the \ebreak\ isosurface).
In other words, their measurements are weighted by mass flux and ours are weighted by surface area.
We expect that this difference in methodology explains why our results indicate that inflow starts much earlier in our runs; early time inflow that doesn't correspond to mass growth won't be picked up by their measurements.
Because our work focused on measuring \vturb, rather than \vinflow, we defer detailed scaling of \vinflow\ to followup work.

\citet{gronke20a} found that cold phase mass evolution's convergence in a $\Mw=6$ simulation run at $\rcl/\cellwidth=8,32$ to be quite poor.
In contrast we found that the cold phase mass evolution in our  $\rcl/\cellwidth=8,16$ runs of our $\Mw=6$ simulation to be fairly well converged.
While it's possible that we could see differences at higher resolution, it's plausible this difference could arise from differences in the cloud temperature.
The clouds in \citet{gronke20a} had a temperature of $\Tcl=4\times 10^4\, {\rm K}$.
This translates to values of \tcoolmin\ and \ecl\ that are factors of $\sim 5$ and $\sim11.7$ larger.
Consequently, we expect $\lturbsonic/\rclcrit$ to be 7.3 times smaller in their simulations, which means they could be under-resolving \rcl\ according to our new resolution criterion.

More generally, one might ask ``How does the choice of \Tcl\ affect our results?'' given that the equilibrium \Tcl\ varies\footnote{The value is commonly controlled by setting a temperature floor or turning off cooling below a certain temperature} greatly among cloud-crushing and galactic outflow studies.
For context, this work focuses on runs with $\Tcl\sim 5\times 10^3\, {\rm K}$, while other works commonly include simulations with $\Tcl\sim 10^4\, {\rm K}$ \citep[e.g.][]{li20a,kanjilal20a,abruzzo22a,schneider20a} or $\Tcl\sim 4\times 10^4\, {\rm K}$ \citep[e.g.][]{gronke18a,gronke20a,abruzzo22a}.
We expect the applicability of our results is more-strongly tied to the shape of $\tcool(T)$ over $\Tcl\la T \la \Tw$ than the precise value of $\Tcl$.
\autoref{fig:vturb-Tcl-dependence} suggests our results are minimally affected when $\tcool(\Tcl)$ exceeds the minimum value of $\tcool$ computed over the temperature range.
However, the applicability is less clear when $\tcool(T)$ is minimized at $\Tcl$ (i.e. if $\Tcl \ga 2\times 10^4\, {\rm K}$ for $p/k_B = 10^3\, {\rm K}\, {\rm cm}^{-3}$, $Z_\odot$, $z=0$) or at a value of $T$ exceeding $\sqrt{\Tcl\Tw}$.
Finally, we note that some works also consider conditions with $\Tcl<500\, {\rm K}$ \citep[e.g.][]{banda-barragan21a, farber22a}.
Further investigation is required to understand the applicability of our results in this context, but our above discussion about $\tcool(T)$'s shape is relevant.

We next draw comparisons with works that studied multiphase gas in turbulent box simulations.
For example, \citet{gronke22a} initialized a pressure-confined cool ($\Tcl=4\times 10^4 {\rm K}$) cloud in a hot ambient background and studied how the system evolved while driving turbulence in the hot phase.
\citet{mohapatra22_filaments} studied the turbulent properties of multiphase gas (comparable to ICM conditions) that emerged from driven turbulence and radiative cooling in a box of initially hot ($T=4\times 10^6 {\rm K}$) gas.
These studies respectively observed that the amplitude of the first and second order velocity structure functions (\vsffirst\ and \vsfsecond) have lower amplitudes in the cold-phase gas than in the other phases, which is in good qualitative agreement with our results.
We note that the sub-Kolmogorv scaling of our \vsfsecond\ measurements are more consistent with the hydrodynamic volume-weighted heating run from \citet{mohapatra22_filaments} than the mass-weighted run.
However, as mentioned in \autoref{sec:vsf}, the driving scale is not sufficiently resolved to remove the bottleneck effect's influence on the slope of \vsfsecond. To be more concrete, we note that \citet{mohapatra22_filaments} illustrated that the driving scale must be resolved by more than 192 cells, in a non-radiative turbulence simulation, to remove the bottleneck effect's influence on the slope.
For that reason, we refrain from making detailed comparisons.

\label{sec:caveats}
\subsection{Caveats}

This work made a number of simplifying assumptions and omitted a variety of potentially relevant physical effects that could potentially modify our results.
Future work should consider:

{\it Other sources of turbulence:} We only analyzed the turbulence that emerged from two phases that initially had coherent velocities without turbulence.
In reality, external processes, like supernovae, can drive turbulence in the wind; this likely alters the interaction's evolution and makes survival more difficult \citep[e.g.][]{schneider20a}.
Additionally, differences in the initial cloud structure, due to turbulent driving before encountering a wind, can affect the rates at which mixing destroy clouds \citep[e.g.][]{schneider17a,banda-barragan19a}.

{\it Thermal Conduction:} The omission of thermal conduction from our simulations will certainly affect the morphology of the cold-phase \citep[e.g.][]{bruggen16a,li20a}.
However, we take solace in the fact that mass transfer through the TRML will be minimally affected in simulations where cooling is fast relative to the mixing time \citep{tan21a}.

{\it Magnetic fields:}
It is well known that magnetic fields can extend the lifetime of clouds \citep[e.g.][]{dusri08a, mccourt15a}.
\citet{banda-barragan18a} showed that magnetic fields have a stabilizing effect on initially turbulent clouds embedded in a laminar wind.
While realistic magnetic field strengths don't seem to strongly affect the criteria for survival through rapid cooling, they do have a number of other effects that will almost certainly affect the system's turbulent properties \citep{gronke20a}.
Among others, such effects include non-thermal support, which could alter cooling properties, suppression of the KH instability and alteration of cloud morphology, leading to higher surface areas \citep{gronke20a}.

{\it Cosmic Rays:}
Cosmic rays were also omitted from our simulations.
They are a known sources of non-thermal pressure support, which may alter cooling properties \citep{butsky20a}.
They can also provide another mechanism for accelerating clouds \citep{wiener19a,huang22a}. 

{\it Gravity:}
Our simulations neglected the effects of gravity because we generally expect our $\chi\leq10^3$ simulations to be Jeans stable.
However, one could imagine that external gravitational fields could sustain an elevated shear velocity \citep{tan23a} and consequently influence the system's turbulent properties.

{\it More realistic cooling:}
All of our simulations assume simplified equilibrium cooling and neglect self-shielding.
However, given that all our simulations where the cloud survives  have $N_{H\textsc{i}}>10^{17.2}\ {\rm cm}^{-2}$, self-shielding may be relevant.
Including more realistic cooling could modify our results \citep{farber22a}, but we leave that for future work.

{\it Viscosity:} Our simulations do not have explicit viscosity \citep{li20a, jennings21a}.
This may affect turbulent properties near the scale of turbulent dissipation.

\section{Conclusion} \label{sec:conclusion}

We have investigated the multiphase turbulent properties that emerge from interactions between cool clouds and hot supersonic flows (or winds).
The relative efficiency of turbulent mixing and radiative cooling in mixing layers govern the outcome of such interactions.
To address difficulties associated with characterizing multiphase turbulence, our analysis employed three distinct methods to measure \vturb.
We found the following primary results for simulations, in which cooling is sufficient for the cloud to survive the interaction and become entrained:

\begin{itemize}
    \item {\bf Radiative cooling dramatically changes the \vturb\ temperature\footnote{For the reader's convenience, we describe phase dependence in terms of temperature even though the majority of this work primarily considers specific internal energy} scaling.}
        In non-radiative simulations \vturb\ has a scaling consistent with the sound speed's temperature scaling: $\vturb\propto\cs\propto\sqrt{T}$.
        In runs with sufficient cooling for entrainment, this scaling only applies for gas colder than \Tbreak, the temperature where \tcool\ is minimized.
        Above \Tbreak, the power-law slope starts near $0.5$ and flattens to ${\sim}0$.
        Consequently, cold gas generally has larger turbulent Mach number and turbulent kinetic energy than hot gas.

    \item {\bf \vturb\ has two stages of temporal evolution.}
        The shear velocity initially drives rapid growth of \vturb\ at early times in the ``pre-entrained'' phase.
        As the cloud becomes partially entrained, \vturb\ drops off before stabilizing at a lower value, one that is of comparable magnitude to the average inflow velocity.

    \item {\bf When comparing different simulations at given points in its evolution, $\vturb(\Tbreak)/\csbreak$ scales with $(\xish\Mw)^{\beta}$ or $((\rcl/\cshot)/\tcoolmin)^\beta$.}
    At early times $\beta\approx1/4$ while at late times $\beta\approx1/2$.

    \item {\bf The driving scale is of order the cloud radius throughout the cloud's entire evolution.}
    
    \item {\bf The grid scale should exceed the minimum cooling length, $\lcool\sim\min(\cs\tcool)$ to resolve 2D phase structure.}
        The 1D temperature phase structure is remarkably well-converged at lower resolutions.

    \item {\bf Our simulations suggest the existence of a minimum length scale for resolving turbulence, \lturbsonic, for clouds with an equilibrium temperature of $5\times 10^3 \la (\Tcl / {\rm K}) \la 2 \times 10^4$.}
        Under-resolving this scale seems to artificially amplify the violence of shattering.
        When this scale is resolved, the entrained cool phase is composed of larger clouds.
\end{itemize}

\begin{acknowledgments}
We thank M. Gronke for useful discussions and for sharing some sample code to compute the velocity structure function.
We are grateful to James Bordner, Mike Norman, and the other \enzoe\ developers.
GLB acknowledges support from the NSF (AST-2108470, XSEDE), a NASA TCAN award, and the Simons Foundation. DBF is supported by the Simons Foundation through the Flatiron Institute.
\end{acknowledgments}

\vspace{5mm}

\software{numpy \citep{harris20a},
          matplotlib \citep{hunter07a},
          yt \citep{turk11a},
          scipy \citep{virtanen20a},
          pandas \citep{mckinney10a}, %
          scikit-image \citep{scikit-image}, %
          fftMPI (\url{http://fftmpi.sandia.gov}), %
          Launcher Utility \citep{wilson14},
          \grackle\ \citep{smith17a},
          \enzoe (\url{http://cello-project.org})
          }

\appendix

\section{Robustness of metrics at early times}\label{appendix:early-time-vturb}

\begin{figure}
  \center
\includegraphics[width = 3.35in]{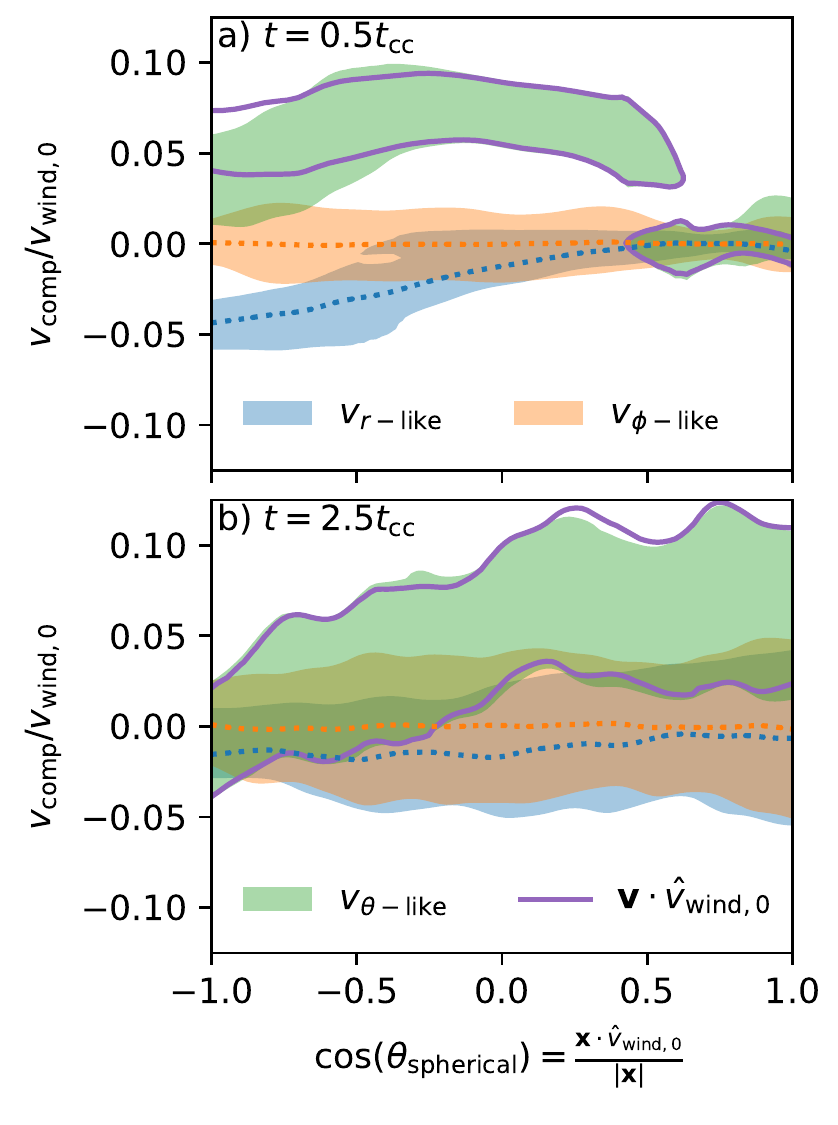}
\caption{\label{fig:spherical-flow}
  The probability density functions of several velocity components (in the cloud's rest-frame), as measured on the $\logXe = 1/6$ iso-surface for our $\chi=1000$, $\xish = 27.8$, $\rcl/\cellwidth=64$ simulation at multiple times.
  The contours bound the region containing the most frequently occurring $68.4\%$ of values at a given $\cos(\theta_{\rm spherical})$.
  The fluctuations in a distribution's mode arises from the mostly-spherical laminar flow at early times.
  The dotted lines show the mean values of \vrlike\ and \vphilike\ as functions of $\cos(\theta_{\rm spherical})$.
  The vertical extent of a contour arises from turbulence (and are somewhat inflated by asymmetries in the flow).
  At early times, estimating \vturb\ from the variance in  any velocity component, other than \vphilike, without explicitly accounting for these laminar variations, will yield over-estimates.
 }
\end{figure}

Our approaches for characterizing \vturb\ all build on the idea that a velocity field can be decomposed into a laminar part and a turbulent part.
Consider an ideal turbulent flow in which the laminar part of the velocity field is uniform.
In this scenario, the magnitude of the laminar part sets the average of the velocity field and the turbulent part sets the dispersion in the velocity values.
For this reason, our methods for measuring a spatially averaged \vturb\ (in a given gas phase) all measure this dispersion in one way or another.

Unfortunately, the flows considered in this work are more complex: the laminar portion of the flow has spatial gradients.
Figure ~\ref{fig:spherical-flow}a illustrates these gradients for several velocity components measured on the $\logXe = 1/6$ iso-surface of our $\chi = 1000$, $\xish = 27.8$ simulation at $0.5\tcc$. 
In more detail, the panel shows the conditional distributions\footnote{These distributions were approximated with kernel density estimation.} of multiple velocity components as a function of $\cos \theta_{\rm spherical}$, where $\theta_{\rm spherical}$ is the polar angle measured from the center of the inflow boundary.

Unless they are removed, such gradients can dominate or inflate the dispersion of the global velocity distribution, which can bias our measurements of \vturb.
\autoref{fig:spherical-flow}b, suggests that this is less of an issue after early times (once \vturb\ has had time to grow) because the dispersion from turbulence is larger relative to the laminar variations.
However, it's clear that these gradients still remain problematic in the wind-aligned velocity component.
\autoref{fig:diff_flow}b shows that large variations in the wind aligned velocity persist to later times, even as the cloud is accelerated.

We expect our \vturb\ measurements from our geometric approach to be unaffected by this issue because it estimates \vturb\ from the dispersion in \vphilike, which maintains a mean of zero at all times.
However, the laminar variations will bias the measurements using our other approaches at early times.
While one might expect our filtering measurements to be resilient to this effect, because it uses a local estimate of the laminar flow, at least some bias will remain given that these early-time gradients are most naturally described in spherical components.
Throughout this work, we elect to just focus on turbulence in velocity components orthogonal to the wind direction, in our filtering and \vsfsecond\ measurements, in order to avoid biases from the wind-aligned velocity component.

As an aside, the resilience of our geometric approach to these biases are related to the definition of the velocity components.
Consider $\hat{u}_{r-{\rm like}}$, which we define the unit vector parallel to the specific internal energy gradient (i.e. $\hat{u}_{r-{\rm like}} = \Vec{\nabla} e/||\Vec{\nabla} e||$).
Because this vector is always normal to the specific internal energy isosurfaces, we can define \vrlike\ and \vphilike\ at arbitrary locations using $\vrlike = -\Vec{v}\cdot \hat{u}_{r-{\rm like}}$ and $\vphilike = \Vec{v} \cdot \left(\vwindhat \times \hat{u}_{r-{\rm like}}\right)$.
Future work may wish to perform filtering or compute the structure function in terms of these components.

\bibliography{ref}{}
\bibliographystyle{aasjournal}

\end{document}